\documentclass[aps,prl,twocolumn,groupedaddress,showkeys,longbibliography]{revtex4-1}

\usepackage{graphicx,amsmath,color}
\usepackage{titlesec}
\usepackage[caption=false]{subfig}

\newcommand\largesection{%
  \titleformat{\section}
    {\normalfont\huge\bfseries\filcenter}{\thesubsection}{1em}{}
}
\newcommand\stdsection{%
  \titleformat{\section}
    {\normalfont\large\bfseries}{\thesubsection}{1em}{}
}

\begin{document}

\title{Spin angular momentum in planar and cylindrical waveguides induced by transverse confinement and intrinsic helicity of guided light}
\author{Diego R. Abujetas} 
\author{Jos\'e A. S\'anchez-Gil}
\email{j.sanchez@csic.es}
\affiliation{Instituto de Estructura de la Materia (IEM-CSIC), Consejo Superior de Investigaciones Cient\'{\i}ficas,\\ Serrano 121, 28006 Madrid, Spain}

\date{\today}

\begin{abstract} 

In recent years, extraordinary spin angular momenta have been investigated in a variety of structured electromagnetic waves, being of especial interest 
in sub-wavelength evanescent fields. Here we demonstrate analytically  that, in planar and cylindrical waveguides supporting transverse electric/magnetic 
modes, transverse spin density arises inside the waveguide (different from the spin induced in the evanescent region outside the waveguide), carrying 
indeed longitudinal extraordinary (so-called) Belinfante's spin momentum. Such contribution depends linearly on the mode transverse wave vector, and is 
thus induced by mode confinement. Cylindrical waveguides  support in addition hybrid modes that exhibit a richer phenomenology with not only azimuthal 
(confinement-related) spin, but also an intrinsic helicity which leads to longitudinal spin density and transverse helicity-dependent spin momentum. 
Results are indeed presented for configurations relevant to spin-orbit coupling in nanophotonic waveguides and to manipulating optical forces in 
IR-to-microwave water-filled channels. Thus guided modes intrinsically carrying confinement-induced transverse spin, combined with intrinsic-helicity-induced 
longitudinal spin (when hybrid), hold promise of superb devices to control spin-orbit interaction and optical forces within confined geometries throughout the 
electromagnetic spectra.

\end{abstract}

\maketitle
\section{Introduction}

Spin and orbital  angular momentum has been a subject of fundamental interest since long ago.  
Recall that  plane waves carry longitudinal orbital angular momentum (OAM) given by its wave vector, and also 
longitudinal spin angular momentum (SAM) arising from its intrinsic helicity due to circular polarization; they 
can be associated to, respectively, spatial and polarization degrees of freedom that can be straightforwardly separated in 
paraxial wavefields. Nonetheless, recent work carried out on SAM and OAM in a variety of optical fields beyond paraxial fields 
\cite{Aiello2015a,Bekshaev2015,Bliokh2015,Bliokh2015b,Gong2018} reveals a wealth of spin-orbit interactions (SOI) 
of light that are attracting a great deal of attention nowadays \cite{Bliokh2015a}.

In this regard, non-paraxial, subwavelength-structured wavefields appear naturally in Plasmonics and Nano-Optics. Therein the vector nature of 
electromagnetic waves has to be fully accounted for, so that spatial and polarization properties can no longer be decoupled, leading to a variety 
of novel phenomenology and functionalities at the nanoscale where spin-orbit interactions  play a crucial role \cite{Bliokh2015a}. 
Particularly relevant and widespread in nano-optics is the emergence of transverse SAM, theoretically described in connection with evanescent 
waves: such polarization-independent, transverse spin component stems from the spatially decaying field of the evanescent wave away from the 
interface, and carries in turn a longitudinal (Belinfante's) SAM  \cite{Bliokh2014a}. Transverse SAM has been later on studied in several geometries
involving well known evanescent waves \cite{Bliokh2014,OConnor2014,Aiello2015a,Bliokh2015,Sayrin2015,VanMechelen2016,Antognozzi2016,Picardi2017,Picardi2018,Neugebauer2018} 
such as plasmons, total internal reflection, etc. In this regard, it is worth mentioning that the following experimental works have exploited 
spin-orbit interaction also in  waveguides \cite{Petersen2014,Mitsch2014,LeFeber2015,Alizadeh2016,Coles2016}, 
in most cases enforcing through the spin-orbit locking the coupling of a circularly polarized exciting field (spin locked in the evanescent tail)
into propagating guided waves with ad-hoc directionality (orbit thus driven).

Nonetheless, the peculiar structured wavefield arising \textit{inside} waveguides has not been addressed in detail up to now.
Recall that cylindrical waveguides support a wealth of transverse and hybrid leaky/guided modes recently shown 
to be crucial in determining the optical properties of semiconductor nanowires in relevant problems such as photoluminescence
\cite{Grzela2014,VanDam2015b,Abujetas2017} and absorption \cite{Paniagua-Dominguez2013,Abujetas2015}. Nanowires and fibers are no doubt extremely 
interesting optical platforms  that hold potential of novel nanophotonic devices, wherein SOI and locking inside could be exploited to manipulate 
i.e. quantum well and dot emission and photoluminescence. Subwavelength waveguides in lower-frequency regimes are amenable to SOI phenomena; 
moreover, optical forces and torques in the IR to GHz domain could be manipulated inside i.e. water-filled waveguides \cite{Andryieuski2015}.

In this work, we study theoretically the spin and orbital angular momenta of confined light inside waveguides. In Sec. II, we  show analytically
that guided modes carry transverse spin density and longitudinal (so-called) Belinfante's spin momentum  inside the waveguide, 
connecting them to relevant magnitudes stemming from light confinement such as energy density and transverse wave vector components. 
This is done for two cases: simplest (planar) waveguide geometry with transverse electric/magnetic modes; and a cylindrical (nanowire) 
geometry (much more involved), which supports not only pure transverse modes but also hybrid modes with intrinsic helicity.
Section III exploits the analytical formulation to explore spin-orbit interactions through the resulting SAM and OAM in
nanophotonic waveguide geometries (semiconductor nanoslabs and nanowires). Water-filled waveguides are considered in Sec. IV to illustrate 
optical forces and torques in the microwave regime. Finally, conclusions are summarized in Sec. V. 

\section{Spin and orbital angular momenta in planar and cylindrical waveguides: Formulation}

\subsection{Transverse guided modes in planar waveguides}

\begin{figure}
\includegraphics[width=0.9\columnwidth]{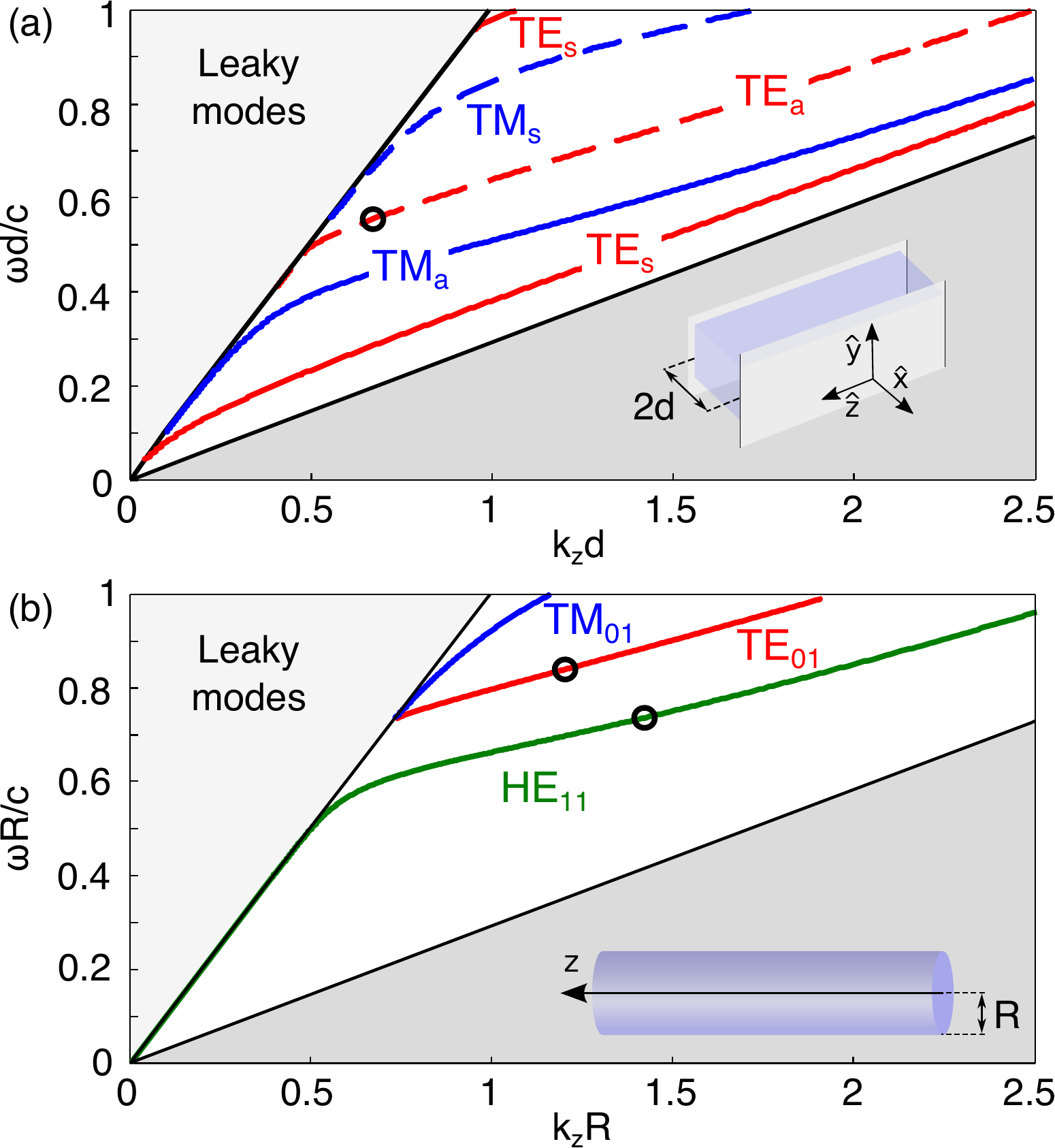}
\caption{ 
(a,b) Dispersion relations for the lowest-order guided modes inside waveguides with refractive index $n=3.43$ (vacuum outside): 
(a) (anti-)symmetric transverse electric and magnetic (TE and TM) guided modes (slab); (b) TE, TM, and hybrid (HE) guided modes (cylinder). 
Circles indicate the specific modes and corresponding wavevectors considered below. 
Insets: Schematic of the dielectric waveguides considered hereafter: (a) slab of width $2d$ and (b) cylinder of radius $R$.}
\label{fig_PW_geo}
\label{fig_NW_geo}
\end{figure}

Let us start with a simple planar waveguide  (see inset in Fig.~\ref{fig_PW_geo}a) consisting of a dielectric slab (core, medium $1$) of thickness $2d$ 
surrounded by another dielectric material (cladding, medium $2$) that has a lower refractive index, $n_1=c\sqrt(\epsilon_1\mu_1)=\sqrt(\epsilon_{r1}\mu_{r1})>n_2$:
we define both relative and absolute dielectric permittivities and magnetic permeabilities ($\epsilon_{r},\mu_{r}$ and $\epsilon,\mu$, respectively) 
for they will be needed below ($c=1/\sqrt{\epsilon_0\mu_0}$ is the speed of light in vacuum). We assume propagation 
along the $z$ direction and translational invariance along the $y$ direction. Transverse electric (TE) and magnetic (TM) modes stand for modes with 
only electric/magnetic field component along the $y$ axis, propagating along the $z$ axis with propagation constant $k_{z}$  and transversal wavevector $k_{t}$. 
For TE modes, the corresponding fields (omitting the time harmonic factor $e^{-\imath\omega t}$) are \cite{Snyder1983}:
\begin{subequations}\begin{eqnarray}
\mathbf{E}&&=\hat{y} \sqrt{\mu_i} A_i f(k_{t}x)e^{\imath k_{z}z}, \\
\mathbf{H}&&= A_i\frac{1}{\omega\sqrt{\mu_i}} [k_{z}f(k_{t}x)\hat{x}+\imath k_{t}f'(k_{t}x)\hat{z}] e^{\imath k_{z}z},
\end{eqnarray}\label{eq_EHTE}\end{subequations}
with $i=1,2$ and 
\begin{subequations}
\begin{eqnarray}
f(x)&&=e^{-\alpha x}, x > d, \\
f(x)&&=\begin{Bmatrix} \sin(k_x x) \\  \cos(k_x x) \end{Bmatrix},  |x|< d,  \\
&&=\mp e^{\alpha x}, x < -d. 
\end{eqnarray}\label{eq_fTE}\end{subequations}
The terms in braces denoting antisymmetric (top) and symmetric (bottom) modes, respectively; $f'(k_{t}x)$ denotes derivative with respect to its argument.
The electromagnetic fields for TM modes can be straightforwardly obtained from the above Eqs.~(\ref{eq_EHTE}) by replacing  
$\mathbf{E},\mathbf{H}\Rightarrow\mathbf{H},-\mathbf{E}$ and $\sqrt{\mu}\Rightarrow\sqrt{\epsilon}$. Field amplitudes 
$A_i$ outside/inside the waveguide are connected through corresponding  boundary conditions.
The wavevector $\mathbf{k}$ components in Cartesian coordinates of such fields depend also on the medium, and are given by:
\begin{align}
\mathbf{k} = (k_{t},0,k_{z}), \nonumber \\ \mathrm{with} \;\;\; k_t=\pm\imath\alpha \;(|x|> d) \;\;\; \mathrm{and} \;\;\; k_t=k_{x} \;(|x|< d),
\label{eq_kt}
\end{align}
where we have used the same notation for the transversal component of the wavevector inside and outside the waveguide, bearing in mind that outside is complex, 
its sign $\imath \alpha$ depending on the considered semi-region. The wavevector components are related by:
\begin{subequations}\begin{eqnarray}
k_{z}^2+k_t^2=\epsilon_{r1}\mu_{r1}\left(\frac{\omega}{c}\right)^2=(n_1\frac{\omega}{c})^2, |x|> d,\\
k_{z}^2-\alpha^2=\epsilon_{r2}\mu_{r2}\left(\frac{\omega}{c}\right)^2=(n_2\frac{\omega}{c})^2, |x|< d;
\end{eqnarray}\label{eq_k}\end{subequations}
where $\omega$ is the angular frequency. 
Since the dielectric waveguide is intended to guide the light, the propagation constant $k_z$ has to be in the range $n_2/n_1<ck_{z}/\omega<n_1$ and
will also depend on mode number and polarization. Indeed, upon imposing boundary conditions, we can obtain the corresponding dispersion relation (cf. e.g.
Ref.~\cite{Snyder1983}) that determines the wavevectors of symmetric and antisymmetric, transverse electric and magnetic (TE and TM) guided modes
(see Fig.\ref{fig_PW_geo}a).

The energy density, defined as:
\begin{equation}
W=\frac{\epsilon}{4}|\mathbf{E}|^2+\frac{\mu}{4}|\mathbf{H}|^2,
\label{eq_W}
\end{equation}
is given for both TE and TM modes by:
\begin{subequations}\begin{eqnarray}
W=&&|A_2|^2 \dfrac{k_{z}^2}{2\omega^2}|f(x)|^2, |x| > d, \\
=&& |A_1|^2 \dfrac{1}{2\omega^2} \left[k_{z}^2|f(x)|^2 +\frac{k_x^2}{2}\right], |x| < d, 
\end{eqnarray}\label{eq_WT}\end{subequations}
Next, the Poynting vector density,
\begin{equation}
\mathbf{P}=\frac{1}{2c^2}\Re\left[\mathbf{E}^*\times\mathbf{H}\right]=\frac{1}{2c^2}\Re\left[\mathbf{E}\times\mathbf{H}^*\right],
\label{eq_P}
\end{equation}
yields:
\begin{eqnarray}
\mathbf{P}=&& \hat{z}|A|^2\dfrac{k_{z}}{2c^2\omega}|f(x)|^2, 
\label{eq_PT}
\end{eqnarray}
for both TE and TM modes.

\begin{figure}
\includegraphics[width=1\columnwidth]{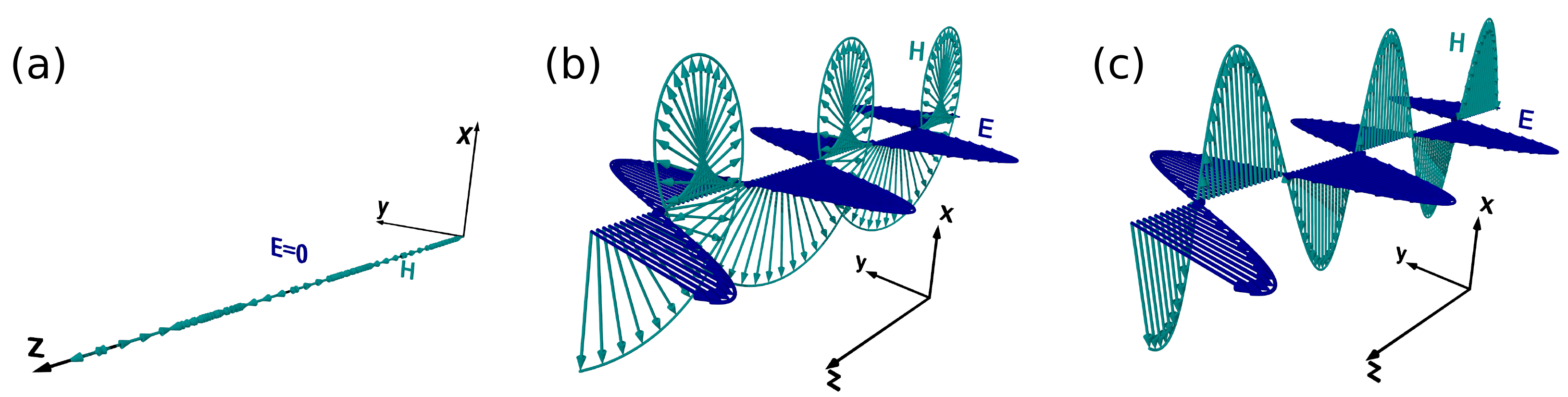}
\caption{ 
Schematic of the electric and magnetic fields along the waveguide from Eqs.~(\ref{eq_EHTE}) and ~(\ref{eq_fTE}) 
for an asymmetric TE mode at transverse positions inside the waveguide satisfying:
(a) $k_xx=0$, (b) $k_xx=\pi/4$, and (c) $k_xx=\pi/2$. TM modes will be identical upon replacing  $\mathbf{E},\mathbf{H}\Rightarrow\mathbf{H},-\mathbf{E}$ . It follows that,
as will be shown below, for TE (respectively, TM) modes, the spin inside the waveguide is $S_e\equiv 0$ (respectively, $S_m\equiv 0$) 
everywhere,  and $S_m\not = 0$ (respectively, $S_e \not = 0$) [as in (b)] except for specific $x$-points where (a) or (c) is satisfied.
}
\label{fig_PW_EH}
\end{figure}

Let us plot the expected electric and magnetic fields along the waveguide from Eqs.~(\ref{eq_EHTE}) and ~(\ref{eq_fTE}), 
for a typical guided mode (asymmetric TE) at given positions inside the waveguide yielding relevant phenomenology.
In Fig.~\ref{fig_PW_EH}a (center of the waveguide for the asymmetric mode), the electric field does vanish, and
the magnetic field has only a nonzero longitudinal component, the fields thus appearing as a longitudinal wave.
In Fig.~\ref{fig_PW_EH}b (corresponding approx. to the center of the lobe of maximum electric field of the lowest-order
asymmetric TE mode), both the electric and magnetic fields have large contributions: the electric one is perfectly transverse,
while the magnetic has a strong longitudinal (out of phase) component, apart from its transverse component. This is no doubt
expected to yield a strong spin. Finally, a case is shown in Fig.~\ref{fig_PW_EH}c where the wave appears purely transverse
(no spin expected). All such varied behavior will certainly give rise to a rich spin phenomenology, as we will now show. 

\subsection{Spin density and momentum in planar waveguides}

The spin density is defined as:
\begin{equation}
\mathbf{S}=\frac{\epsilon_0\mu_r^{-1}}{4\omega}\Im\left[\mathbf{E^*}\times\mathbf{E}\right]+\frac{\mu_0\epsilon_r^{-1}}
{4\omega}\Im\left[\mathbf{H^*}\times\mathbf{H}\right]=\mathbf{S}_e+\mathbf{S}_m.
\label{eq_s}
\end{equation}
Recall that the helicity, 
\begin{eqnarray}
h=&& - \frac{\sqrt{\epsilon_{0}\mu_{0}}}{2\omega}\Im[\mathbf{E^*}\cdot\mathbf{H}],
\label{eq_h}
\end{eqnarray}
of all guided modes in this planar waveguide is $h=0$.

From the electric and magnetic field components, Eqs. (\ref{eq_EHTE}) and related, it follows that
$\mathbf{S}_e=0$ ($\mathbf{S}_m=0$) for TE (respectively, TM) guided modes. The only nonzero components of the spin densities are 
$\mathbf{S}_m=\hat{y}S_m$ for TE modes and $\mathbf{S}_e=\hat{y}S_e$ for TM modes. 
Upon introducing them into Eq. (\ref{eq_s}), we end up with the resulting total spin density:
\begin{subequations}\begin{eqnarray}
\mathbf{S}=\hat{y}&& |A_2|^2\dfrac{k_{z}\alpha}{2\omega^3n_2^2} e^{-2\alpha x}, x > d,  \\
=\hat{y}&& \begin{Bmatrix} - \\ + \end{Bmatrix} |A_1|^2 \dfrac{k_{z}k_x}{4\omega^3n_{1}^2}  \sin(2k_x x), |x| < d,\label{eq_sTEM_in} \\
=-\hat{y}&&|A_2|^2\dfrac{k_{z}\alpha}{2\omega^3n_2^2} e^{2\alpha x}, x < -d;
\end{eqnarray}\label{eq_sTEM}\end{subequations}
for both TE and TM modes, where we have made use of $\epsilon_r\mu_r=n^2$. Note that it can be written in a compact form as follows:
\begin{eqnarray}
\mathbf{S}=\hat{y}&& |A_2|^2\dfrac{k_{z}k_{t}}{4\omega^3n_2^2} \dfrac{d}{d(k_{t}x)}|f(k_{t}x)|^2. 
\end{eqnarray}
Three features should be emphasized at this point regarding the spin density above (\ref{eq_sTEM}): (i) unlike that of the evanescent field, it can
be locally positive or negative inside the waveguide [cf. (\ref{eq_sTEM})]; (ii) it is either purely magnetic ($S_e=0$) or electric ($S_m=0$)
for, respectively, TE or TM guided modes, and (iii) it is proportional to the transverse component of the wavevector $k_{t}$.

We now turn to calculate the extraordinary spin momentum, obeying:
\begin{equation}
\mathbf{P}^S=\frac{1}{2}\nabla\times\mathbf{S}.
\label{eq_ps}
\end{equation} 
Since the spin density has only a nonzero component along the $y$ axis [cf. Eqs. (\ref{eq_sTEM})], which in turn 
does not depend on the propagation direction $z$, the  spin momentum has only a nonzero component along $z$, namely:
\begin{eqnarray}
\mathbf{P}^S =\hat{z}&& |A|^2\dfrac{k_{z}k_{t}^{2}}{8\omega^3n^2} \dfrac{d^{2}}{d(k_{t}x)^{2}}|f(k_{t}x)|^2. 
\label{eq_psTEM}
\end{eqnarray}
for both TE and TM modes. As the spin density, the spin momentum is either purely magnetic (TE modes) or electric (TM modes). 

Finally, the canonical (orbital) part $\mathbf{P}^O$ of the momentum density can be simply obtained from:
\begin{equation}
\mathbf{P}=\mathbf{P}^O+\mathbf{P}^S.
\label{eq_po}
\end{equation}
Therefore, from Eqs. (\ref{eq_PT})  and  (\ref{eq_psTEM}), it follows that:
\begin{subequations}\begin{eqnarray}
\mathbf{P}^O= \hat{z}&& 
|A_2|^2\dfrac{k_{z}^3}{2n_2^2\omega^3} e^{\mp 2\alpha x}, \;\;|x| > d,  \\
=\hat{z}&& |A_1|^2 \dfrac{k_{z}}{2n_{2}^2\omega^3} \left[k_{z}^2|f(x)|^2+\frac{k_x^2}{2}\right], |x| < d, 
\end{eqnarray}\label{eq_PO}\end{subequations}
for both TE and TM modes. Incidentally, we have verified that the same $\mathbf{P}^O$ is obtained if directly calculated from its canonical expression \cite{Bliokh2015b}.
Nonetheless, the orbital momentum presents both contributions, electric and magnetic. It follows from the latter equations that the orbital momentum 
in both polarizations is proportional to the wavevector along the propagation direction in the expected manner:
\begin{eqnarray}
\frac{P^O_z}{W}=\frac{k_{z}}{\omega n^2}.
 \label{eq_poW}
\end{eqnarray}In addition, it can be shown that the spin momentum can be expressed as:
\begin{align}
\frac{P^S_z}{W} = \dfrac{k_{t}^2}{\omega n^2k_{z}} [1-\tilde{P}_{NP}(\mathbf{r})],
\label{eq_PSW_NP}
\end{align}
where $\tilde{P}_{NP}(\mathbf{r})$ is a non-paraxial term (vanishes for paraxial waves),
\begin{align}
\tilde{P}_{NP}(\mathbf{r}) = \dfrac{1}{8k_{t}^{2}} \left\lbrace (\mathbf{k}\cdot\mathbf{E})(\mathbf{k}\cdot\mathbf{E}^{*})\epsilon + (\mathbf{k}\cdot\mathbf{H})(\mathbf{k}\cdot\mathbf{H}^{*})\mu \right. \nonumber \\
+ \left[ \mathbf{k}\times \mathbf{E} - \mu\omega \mathbf{H} \right]\left[ \mathbf{k}\times \mathbf{E}^{*} - \mu\omega \mathbf{H}^{*} \right]\epsilon \nonumber \\
\left. + \left[ \mathbf{k}\times \mathbf{H} + \epsilon\omega \mathbf{E} \right]\left[ \mathbf{k}\times \mathbf{H}^{*} + \epsilon\omega \mathbf{E}^{*} \right]\mu \right\rbrace.
\label{eq_nP}
\end{align}
Therefore, it is evident from Eqs.~(\ref{eq_sTEM_in}) and~(\ref{eq_psTEM}), the main results of this subsection, that a transverse SAM arises inside 
planar waveguides, which in turn yields an extraordinary longitudinal spin momentum, both proportional to the guided mode transverse wavevector component. The
contribution from the longitudinal spin momentum, which, unlike in the evanescent region, can be positive or negative inside the waveguide (as we will show below),
is indeed crucial to retrieve the proper dependence of the canonical momentum on mode wavevector~(\ref{eq_poW}).
Before discussing in detail these terms, we will show next that similar transverse SAM arise in a more complex waveguide geometry in order to 
assess the universal character in connection to guided light.

\subsection{Transverse and hybrid guided modes in cylindrical waveguides}

Let us now study cylindrical waveguides, which support a wealth of guided modes exhibiting a rich spin phenomenology. The fields of the waveguide can be expressed in 
cylindrical coordinates [see Fig.~\ref{fig_NW_geo}(b)] as follows:
\begin{subequations}\begin{align}
& E_{r} = \sum{\left[ \frac{\imath k_{z}}{k_{t}} Z_{m}^{'} \left( k_{t} r \right) a_{m} - \frac{\mu \omega m}{k_{t}^{2} r} Z_{m} \left( k_{t} r \right) b_{m} \right] F_{m}}, \\
& E_{\theta} = - \sum{\left[ \frac{mk_{z}}{k_{t}^{2} r} Z_{m} \left( k_{t} r \right) a_{m} + \frac{\imath \mu \omega}{k_{t}} Z_{m}^{'} \left( k_{t} r \right) b_{m} \right] F_{m}}, \\
& E_{z} = \sum{\left[ Z_{m} \left( k_{t} r \right) a_{m} \right] F_{m}}, \\
& H_{r} = \sum{\left[ \frac{m k^{2}}{\mu \omega k_{t}^{2} r} Z_{m} \left( k_{t} r \right) a_{m} + \frac{\imath k_{z}}{k_{t}} Z_{m}^{'} \left( k_{t} r \right) b_{m} \right] F_{m}}, \\
& H_{\theta} = \sum{\left[ \frac{\imath k^{2}}{\mu \omega k_{t}} Z_{m}^{'} \left( k_{t} r \right) a_{m} - \frac{mk_{z}}{k_{t}^{2} r} Z_{m} \left( k_{t} r \right) b_{m} \right] F_{m}}, \\
& H_{z} = \sum{\left[ Z_{m} \left( k_{t} r \right) b_{m} \right] F_{m}},
\end{align}\label{eq_EHNW}\end{subequations}
where $k_{z}$ and $k_{t}$ are again the longitudinal and transverse component component of the wavevector, respectively, and $k = |\mathbf{k}|$ is the modulus of the wavevector. 
Recall that $k_{t}$ is imaginary outside the waveguide. The function $Z_m\left( k_{t} r \right)$ is the appropriate Bessel function 
that matches the boundary conditions \cite{Paniagua-Dominguez2013,Snyder1983} and $Z'_m\left( k_{t} r \right) $ denotes the derivative with respect to the argument. Upon using functions $Z_m$, we collapse 
EM fields inside and outside the waveguide into formally identical equations. In this regard, recall that $a_m,b_m$, ..., obey different expressions inside/outside. 
The function $F_{m}= e^{\imath \left( m\theta +k_{z}-\omega t \right)}$ is the phase of the wave, the subscript $m$ being an integer related to its azimuthal 
order. Bear in mind that the longitudinal and transverse components of the wavevector depend on $m$, but we dropped the subscript for simplicity. For guided modes, 
the wavevector components are defined as in Eq.~(\ref{eq_k}) (assuming vacuum outside $\epsilon_r=\mu_r=1$), replacing the transverse components $k_x$ and $\imath\alpha$ 
in Eq.~(\ref{eq_kt}) by its transversal (radial) $k_t$ and $\imath\alpha$ counterparts associated with the cylindrical geometry.

The fields of the waveguide written in Eq~(\ref{eq_EHNW}) are expressed as the sum over different guided modes, encoded in the subindex $m$. 
The guided modes satisfy the corresponding dispersion relation (not shown here, cf. Refs. \cite{Paniagua-Dominguez2013,Abujetas2015,Snyder1983});
solutions can be associated to each guided mode, labeled by a pair index $ml$, where $m=0,\pm 1,\pm 2,\ldots$ is the azimuthal index and  $l=1,2,3,\ldots$ the radial index.
For guided modes with $m=0$, the field is symmetric about the axis, exhibiting a pure transverse character, either electric (TE$_{0l}$, $a_m=0$ so that 
$E_r = E_z = H_{\phi}=0$) or magnetic (TM$_{0l}$, $b_m=0$ so that $H_r = H_z = E_{\phi}=0$). Hybrid modes arise for $m\neq 0$ (HE$_{ml}$). We show in 
Fig.~\ref{fig_NW_geo}(b) the dispersion relation of the lowest-order modes: TM$_{01}$, TE$_{01}$, and  the hybrid HE$_{11}$ (no cutoff, lowest-order mode), 
which are the guided modes we will considered explicitly below.

We now proceed to calculate the generic expressions for the  energy density,  Poynting vector, and helicity,  from Eqs.~(\ref{eq_W}), (\ref{eq_P}), 
and~(\ref{eq_h}), for arbitrary guided modes in a cylindrical lossless waveguide with electromagnetic fields given by Eqs.~(\ref{eq_EHNW}). 
This is done in the Supplemental Material \cite{Supp}.
In addition, we also show therein
the generic expressions for all spin-related magnitudes, namely:
spin density and momentum, Eqs.~(\ref{eq_s}) and~(\ref{eq_ps}), and resulting orbital momentum~(\ref{eq_po}). 
Interestingly, we should emphasize that the orbital momentum $P^O$ satisfies the expected dependence on longitudinal wavevector $k_{z}$ as follows: 
\begin{equation}
\dfrac{P^O}{W}=\dfrac{k_{z}}{\omega n^2}.
\label{eq_POWNW}
\end{equation}
Let us now discuss all these relevant magnitudes for given lowest-order guided modes with most relevant symmetries.

\subsection{Transverse guided modes: Confinement-induced SAM}

In the case of transverse guided modes ($m=0$), it is evident from Eq.~(S6) 
that the helicity vanishes ($h=0$). Thus the spin density should vanish  
except for the evanescent component of the EM fields outside the waveguide. However, it follows from Eq.~(S8), 
for the specific case of TE modes, 
that the spin density vanishes neither outside (as expected) nor inside the waveguide:
\begin{subequations}\begin{eqnarray}
&& S_{r}  =S_{z} = 0,  \\
&& S_{\theta} = \dfrac{\mu k_{z}}{2n^{2}\omega k_{t}} \left|b_{0}\right|^{2} Z_{0}^{*} \left( k_{t} r \right) Z_{0}^{'}\left( k_{t} r \right);   
\end{eqnarray}\label{eq_SNW_TX}\end{subequations}
which leads to an  extraordinary spin momentum with a longitudinal component, as follows:
\begin{equation}
P_{z}^{S} = \dfrac{\mu k_{z}}{4n^{2}\omega}  |b_{0}|^{2} \left[ \dfrac{k_{t}^{2}}{|k_{t} |^{2}}\left|Z_{0}^{'} \left( k_{t} r \right)\right|^{2}  
- \left|Z_{0} \left( k_{t} r \right)\right|^{2}\right].
\label{eq_PSNW_TX}\end{equation}
The electromagnetic properties for the TM modes follow the same expressions as for TE waves after replacing $|b_{0}|^{2}\mu \rightarrow |a_{0}|^{2}\epsilon$. 
Importantly, the contribution to the spin density and momentum for TE (TM)  waves is fully magnetic (electric). Actually, if we revisit  Eqs.~(\ref{eq_EHNW}) for 
TE guided modes, we realize that, at given transverse positions within the cylindrical waveguide, the electric and magnetic fields exhibit a behavior analogue to that 
shown in Fig.~\ref{fig_PW_EH} for planar waveguides, with an imaginary longitudinal component of $\mathrm{H}$ leading to the magnetic spin density (for TM modes, 
it is the complex $\mathrm{E}$ which yields the electric spin density).

Moreover, after rewriting the energy density, Eq.~(S2), 
for pure transverse modes as:
\begin{equation}
W = \dfrac{1}{4}|b_{0}|^{2}\mu \left[\dfrac{ k_{z}^{2} + k^{2}  }{|k_{t} |^{2}}\left|Z_{0}^{'} \left( k_{t} r \right)\right|^{2} +  \left|Z_{n} \left( k_{t} r \right)\right|^{2} \right],
\label{eq_WNW_TX}
\end{equation}
Remarkable, the spin momentum satisfies the same expression as for planar waveguides:
\begin{align}
\frac{P^S_z}{W} = \dfrac{k_{t}^2}{\omega n^2k_{z}} [1-\tilde{P}_{NP}(\mathbf{r})],
\label{eq_PSWNW_NP}
\end{align}
with the non-paraxial term $\tilde{P}_{NP}$ given by Eq.~(\ref{eq_PSW_NP}). Recall that, for  weakly guided waveguides, 
the electromagnetic field inside becomes paraxial \cite{Snyder1983}; this confirms the crucial role of confinement in the 
emergence of large spin density and momentum inside waveguides.

Therefore, the transverse spin density and longitudinal spin momentum inside both planar and cylindrical waveguides do not vanish 
(despite not being evanescent) and depend on the transverse momentum $k_{t}$; whereas the orbital momentum, Eq.~(S13), 
has been shown above to depend as expected on the guided mode wavevector along the propagation direction $k_{z}$. 
Incidentally,   transverse momentum has been  associated to an  effective mass $k_t\sim m$ if we rewrite Eq.~(\ref{eq_k}) as \cite{Wang2007b,Zang2016} 
$\omega= \sqrt{k_{t}^2+k_{z}^2}\propto\sqrt{m^2+p^2}$, which thus formally underlies the emergence of extraordinary SAM for guided modes.

\subsection{Hybrid guided modes: Intrinsic-helicity-induced spin}

We now turn to study the spin angular momentum of the HE$_{11}$ hybrid guided mode, which in fact exhibit strong intrinsic helicity given by:
\begin{align}
& h = \dfrac{1}{2 \omega n} \left\lbrace \dfrac{2k_{z} k}{|k_{t} |^{2}k_{t} r}  Z_{1} \left( k_{t} r \right) Z_{1}^{'*}\left( k_{t} r \right)\left[|a_{1}|^{2}\epsilon + |b_{1}|^{2}\mu \right] \right. \nonumber \\
&\left. + \left[ \left|Z_{1} \left( k_{t} r \right)\right|^{2} + \left( \left|Z_{1}^{'} \left( k_{t} r \right)\right|^{2}  + \dfrac{ \left|Z_{1} \left( k_{t} r \right)\right|^{2}}{|k_{t} |^{2} r^{2}} \right) \dfrac{k_{z}^{2} + k^{2}}{|k_{t} |^{2}} \right] \right. \nonumber \\
&\hspace{1in}\left. \times \Im\left[a_{1}b_{1}^{*} \right]\sqrt{\epsilon\mu} \right\rbrace ;
\label{eq_hNW_HE}
\end{align}
the corresponding energy density from Eq.~(\ref{eq_W}) is:
\begin{eqnarray}
 && W = \dfrac{1}{4} \left\lbrace\left[ \left|Z_{1} \left( k_{t} r \right)\right|^{2} + \left( \left|Z_{1}^{'} \left( k_{t} r \right)\right|^{2}  + \dfrac{ \left|Z_{1} \left( k_{t} r \right)\right|^{2}}{|k_{t} |^{2} r^{2}} \right)
\right.\right. \nonumber \\  
&& \hspace{1in} \left.\times \dfrac{k_{z}^{2} + k^{2}}{|k_{t} |^{2}} \right]  \left[|a_{1}|^{2}\epsilon + |b_{1}|^{2}\mu \right]  \nonumber \\ 
&& \hspace{.5in}\left.+ \dfrac{8\sqrt{\epsilon \mu} }{|k_{t} |^{2}} \dfrac{k_{z}k}{k_{t} r} Z_{1} \left( k_{t} r \right) Z_{1}^{'*}\left( k_{t} r \right) \Im\left[a_{1}b_{1}^{*} \right]\right\rbrace.
\label{eq_WNW_HE}
\end{eqnarray}
 We can write Eq.~(S8), 
 which becomes in this case very involved, as follows:
\begin{align}
&S_{r} =  0, \\
&S_{\theta}= \dfrac{1}{2\omega\epsilon_{r}\mu_{r}}  \left[ \dfrac{k_{z}}{k_{t}} Z_{1}^{*} \left( k_{t} r \right) Z_{1}^{'}\left( k_{t} r \right)\left[|a_{1}|^{2}\epsilon + |b_{1}|^{2}\mu \right] \right. \nonumber \\
&\hspace{1cm} \left. + \dfrac{2k}{k_{t}^{2}r} \left|Z_{1} \left( k_{t} r \right)\right|^{2} \Im\left[a_{1}b_{1}^{*} \right] \sqrt{\epsilon \mu} \right],  \\
& S_{z} = \dfrac{|k_{t} |^{-2}}{2\omega\epsilon_{r}\mu_{r}}\left\lbrace  \dfrac{k_{z}^{2} + k^{2} }{k_{t}r}Z_{1} \left( k_{t} r \right) Z_{1}^{'*}\left( k_{t} r\right) 
\left[|a_{1}|^{2}\epsilon + |b_{1}|^{2}\mu \right]  \right. \nonumber \\  
& \hspace{1cm}\left. +2k_{z}k  \left[ \left|Z_{1}^{'} \left( k_{t} r \right)\right|^{2}  + \dfrac{ \left|Z_{1} \left( k_{t} r \right)\right|^{2}}{|k_{t} |^{2} r^{2}}\right]
\Im\left[a_{1}b_{1}^{*} \right]\sqrt{\epsilon \mu} \right\rbrace.
\label{eq_SNW_HE}\end{align}
Now it is actually not trivial to relate the spin density from Eqs.~(S8) 
either to the helicity~(\ref{eq_hNW_HE}) or to the energy density (\ref{eq_WNW_HE}). 
However, it is evident from Eqs. ~(\ref{eq_SNW_HE}), that: (i) the transverse spin density $S_{\theta}$  includes both electric and magnetic contributions (first term) 
along with a new hybrid  contribution (second term); (ii) the spin density yields a longitudinal contribution $S_z$ stemming from intrinsic helicity. 
The corresponding expressions for the orbital and spin momenta ($P_{z}^{O}$, $P_{\theta}^{O}$, $P_{z}^{S}$ and $P_{\theta}^{S}$) are given in the Supplemental Material \cite{Supp}. 
Interestingly, the longitudinal components of the orbital and the spin angular momentum can be related to the energy as follows:
\begin{equation}
\dfrac{P_{z}^{O}}{W} = \dfrac{k_{z}}{\omega n^{2}}, \quad \dfrac{P_{z}^{S}}{W} = \dfrac{k_{t}^{2}}{n^{2}k_{z}\omega } \left[1-\tilde{P}_{NP}\right],
\label{eq_POWNW_HE}\end{equation}
as in the case of pure transverse modes shown above.
\begin{figure}
\includegraphics[width=0.9\columnwidth]{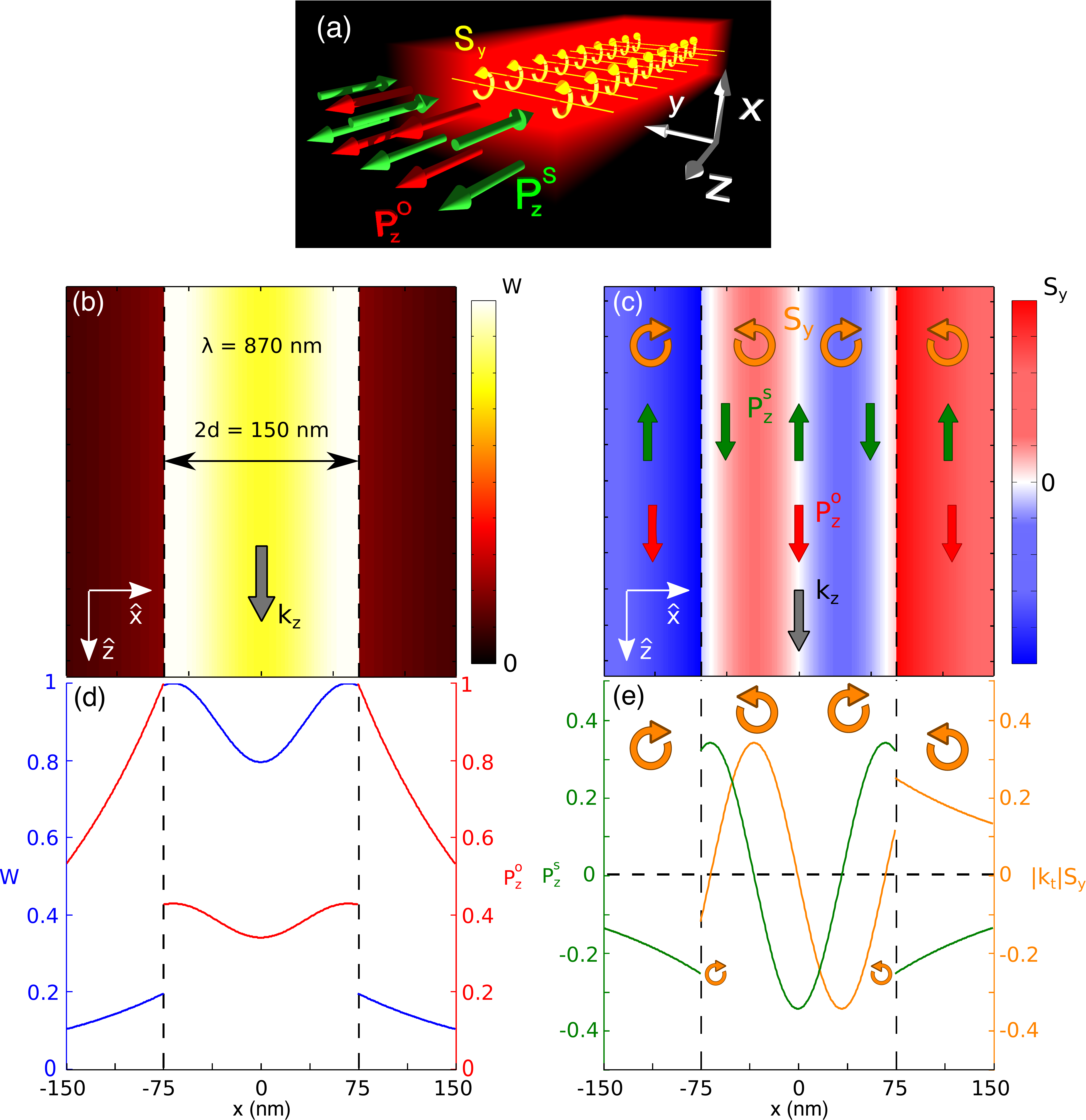}
\caption{(a) Schematic depicting the orientation of orbital $\mathbf{P}^O$ (red arrows) and    spin 
$\mathbf{P}^S$ (green arrows)  momenta, inside the planar waveguides, with loops inside illustrating the 
confinement-induced spin density $\mathbf{s}$.  (b-e) The anti-symmetric lowest-order TE mode is considered
for a planar waveguide of width $2d$ and $\epsilon_r=11.76$ for normalized  half-width $\omega d/c$=0.54 
(which corresponds to an InP nanoslab with thickness $2d=150$ nm at $\lambda=870$ nm).  
(b,d) Contour map and radial dependence of the energy density $W$, the latter  including also the orbital $P_z^O$ momentum. 
(c,e) Contour map and radial dependence of the only non-zero component of the spin density $S_y$, 
the latter (e) multiplied by the transverse wavevector component and including also the only non-zero component 
of the spin angular momentum $P_z^S$ (normalized by $P_z^O$).}
\label{fig_PW}
\end{figure}

\section{Spin-orbit interactions inside nanophotonic waveguides}

To shed light onto the emergence of confinement-induced spin angular momentum, 
let us plot all relevant magnitudes for planar and cylindrical waveguides with $\epsilon_r=11.76$. Incidentally, this choice of refractive 
index makes our results applicable throughout the visible and near-infrared to many semiconductors with similar refractive index 
\cite{Grzela2014,Abujetas2017,Paniagua-Dominguez2013,Sell2016}, such as crystalline Si, GaP, GaAs, InP, etc.; indeed, it  corresponds approximately 
to the photoluminescence band of InP at $\lambda\sim 870$ nm, which  could be illustrative of SOI in emission processes inside InP nanostructures,
such as slabs or nanowires \cite{Grzela2014}, especially if circularly polarized (or even more complex) dipole sources could be engineered \cite{Picardi2018}.
In fact, circularly polarized luminescence has been recently reported e.g. from coupled InGaN/GaN quantum well and quantum dots structure \cite{Yu2016},
from perovskite nanocrystals suitable to deploy flexible devices \cite {Shi2018a}, also from  colloidal CdS quantum dots \cite{Naito2010,Huo2017}. The latter
colloidal quantum dots could be exploited in micron-sized, liquid-filled waveguides, in which the liquid itself might play the role of the denser optical medium 
(its refractive index needn't be very high indeed). On the other hand, recall that magnetic dipole emission is also available through lanthanide-doped 
nanostructures and nanoparticles \cite{Karaveli2011,Taminiau2012,Liu2013k}.

First, we show in Fig.~\ref{fig_PW} the simplest case: a semiconductor nanoslab supporting the first-order  anti-symmetric TE mode (see the mode dispersion 
relation in Fig.~\ref{fig_PW_geo}). The transverse energy density $W$ is plotted for the sake of comparison in Figs.~\ref{fig_PW}b (transverse color map) 
and~\ref{fig_PW}d; in the latter the orbital momentum is also included to explicitly show that the expected direct proportionality $P^O_z/W= k_z/(\omega n^2)$ 
is satisfied [Eq.~(\ref{eq_poW}), note that the factor $n$ is different inside/outside]. Transverse spin momentum, Eq.~(\ref{eq_sTEM_in}), arises inside the waveguide
with a rich phenomenology depending on the mode symmetry, rotating differently on each half waveguide (evident from the sign of $S_y$ in Fig.~\ref{fig_PW}c,e 
and depicted also by loops). In turn, it results in a corresponding longitudinal spin momentum [Eq.~(\ref{eq_psTEM})] which can point along (or opposite) to the 
guided mode propagation direction, and thus to the orbital angular momentum (evident from the sign of $P^S_z$ and depicted by arrows). Moreover, it is simultaneously 
parallel and antiparallel on each half waveguide.  In fact, the latter behavior can be understood in light of the complex electric and magnetic fields plotted in 
Fig.~\ref{fig_PW_EH}. The spin density maxima occur at symmetric positions inside the waveguide where electric and magnetic fields rotate as in Fig.~\ref{fig_PW_EH}b. 
At the waveguide center, the spin density vanishes as a result of the vanishing of the electric field (see  Fig.~\ref{fig_PW_EH}b); the spin density also vanishes 
near the waveguide boundaries, at positions where the electric and magnetic fields are locally paraxial as in Fig.~\ref{fig_PW_EH}c.

Finally, note that outside the waveguide, the transverse spin density, Eq.~(\ref{eq_sTEM}), stems from the evanescent character of the guided  modes outside, as expected 
from the case of pure evanescent waves \cite{Bliokh2014a}; however, its associated  longitudinal spin momentum [Eq.~(\ref{eq_psTEM})] outside the waveguide (unlike inside)  
always points opposite to the orbital momentum direction. Analogous results for TM modes are included in the Supplemental Material (see Fig. S1 \cite{Supp}) that confirm the emergence of 
transverse spin density with extraordinary longitudinal momentum; as expected, since the TM mode shown in Fig. S1 is more confined than the TE mode shown in Fig.~\ref{fig_PW}, 
the evanescent SAM reaches larger values but decays more abruptly away from the waveguide. The behavior for symmetric modes (not shown here) is very similar, except for the 
fact that spin density/momentum is symmetric/antisymmetric instead. Finally, higher-order guided modes do preserve the symmetry and polarization of the spin-related 
magnitudes discussed above, introducing additional number of alternating-sign layers  (see e.g. Fig.~\ref{fig_GHz} below).
\begin{figure}
\includegraphics[width=0.9\columnwidth]{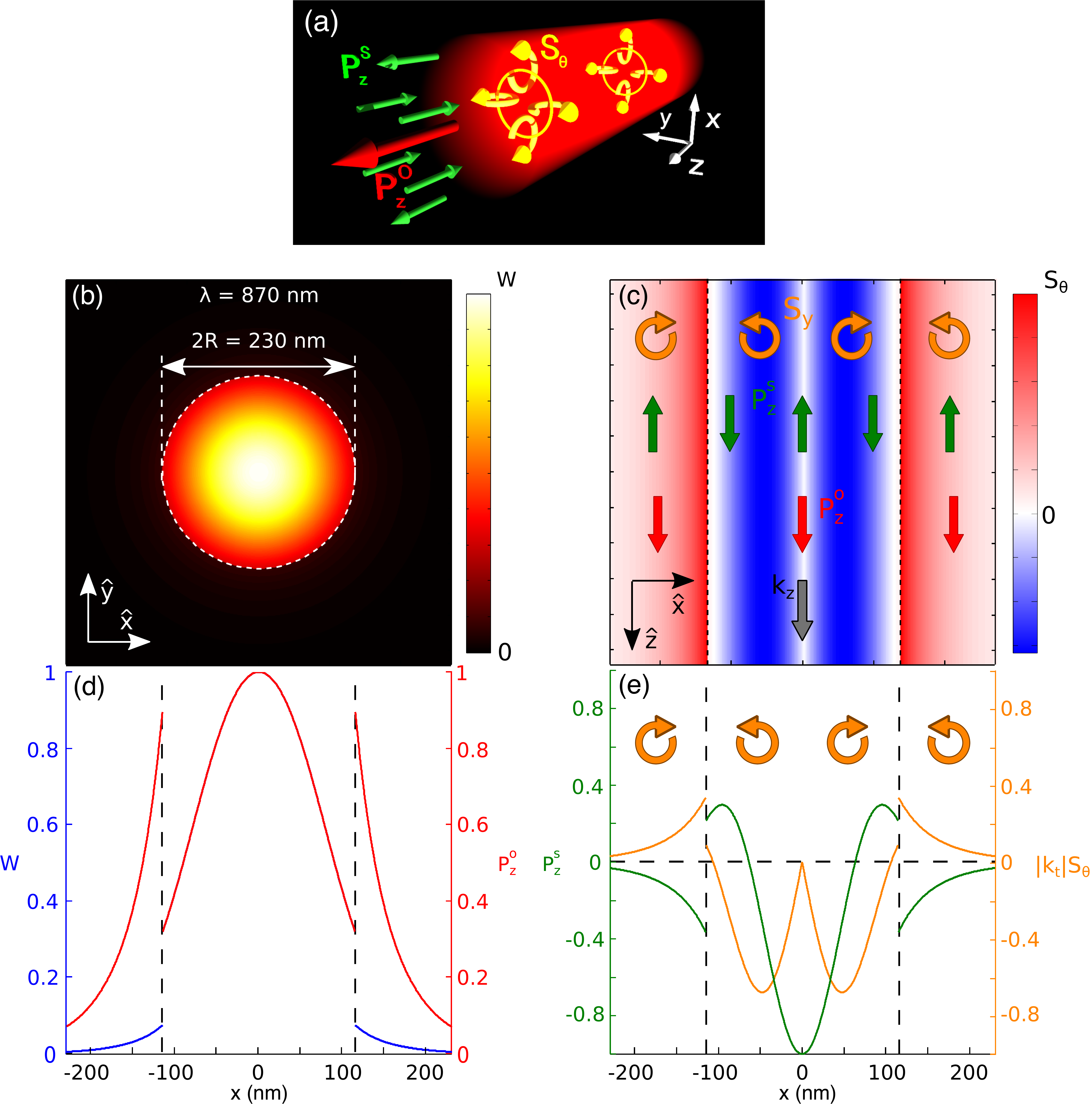}
\caption{(a) Schematic depicting the orientation of orbital (red arrows) and  spin (green arrows)  momenta, along with spin density (loops) inside the cylindrical 
waveguides; as in Fig.\protect{\ref{fig_PW}}. (b-e) The lowest-order TE$_{01}$ guided mode for a cylindrical waveguide of radius $R$ and $\epsilon_r=11.76$ 
is considered for a normalized radius $\omega R/c=0.83$ (which corresponds to an InP nanowire with $2R=230$ nm at $\lambda=870$ nm): 
(b,d) Contour map and radial dependence of the energy density $W$, the latter including also the orbital $P_z^O$ momentum. 
(c,e) Contour map and radial dependence of the only non-zero component of the spin density $S_{\theta}$, 
the latter (e) multiplied by the transverse wavevector component and including also the only non-zero component 
of the spin angular momentum $P_z^S$ (normalized by $P_z^O$).
}
\label{fig_NW_TX}
\end{figure}

Next, let us plot all spin-related magnitudes for the lowest-order TE guided mode (zero helicity) in a cross section of the cylindrical waveguide with 
$\epsilon_r=11.76$ in Fig.~\ref{fig_NW_TX}. First of all, a 3D schematic illustrating the orientation of the corresponding spin (loops) and
momenta (arrows) is depicted in Fig.~\ref{fig_NW_TX}a, clearly revealing its transverse character with axial symmetry. The energy density $W$ is included for the sake of completeness 
in Fig.~\ref{fig_NW_TX}b, revealing the strong confinement of this TE$_{01}$ guided mode for the choice of parameters. Its radial dependence is explicitly plotted in
Fig.~\ref{fig_NW_TX}d, along with the orbital angular momentum, to confirm again its linear dependence $P^O_z/W\sim k_z/(\omega n^2)$, Eq.~(\ref{eq_POWNW}).
The emergence of the transverse confinement-induced  SAM becomes evident in the right column. First, a  strong contribution within the cylinder to the azimuthal spin density 
[electric/magnetic for TE/TM modes, cf. Eq.~(\ref{eq_SNW_TX})], is observed in the color map in Fig.~\ref{fig_NW_TX}c,  which rotates about the cylinder axis. Its radial 
dependence is shown explicitly in Fig.~\ref{fig_NW_TX}e: note that it is zero in the center and close to the boundary, achieving its maximum within a ring inside the cylinder.  
Such behavior can be fully understand in light of the complex electric and magnetic fields represented in Fig.~\ref{fig_PW_EH}, also discussed in connection to Fig.~\ref{fig_PW}, 
bearing in mind that the translational invariance along the transverse direction in the planar waveguide is replaced by the axial symmetry of this cylindrical waveguide.
The corresponding  extraordinary longitudinal spin momentum [Eq.~(\ref{eq_PSNW_TX})] is also strong inside the waveguide, concentrated at the center and pointing along the 
cylinder axis (see Fig.~\ref{fig_NW_TX}e), opposite to the guided mode propagation direction, except for a thin corona near the exterior boundary where it is parallel.  
A weaker transverse spin density  is also observed in Fig.~\ref{fig_NW_TX}c and.~\ref{fig_NW_TX}e, which stems from the evanescent character of the guided  modes outside 
[Eq.~(\ref{eq_SNW_TX})]. Indeed, the spin rotation and momentum direction outside (antiparallel with respect to orbital momentum), unlike those inside, is also  fixed, 
similarly to what was observed in planar waveguides (see Fig.~\ref{fig_PW}).
\begin{figure}
\includegraphics[width=0.9\columnwidth]{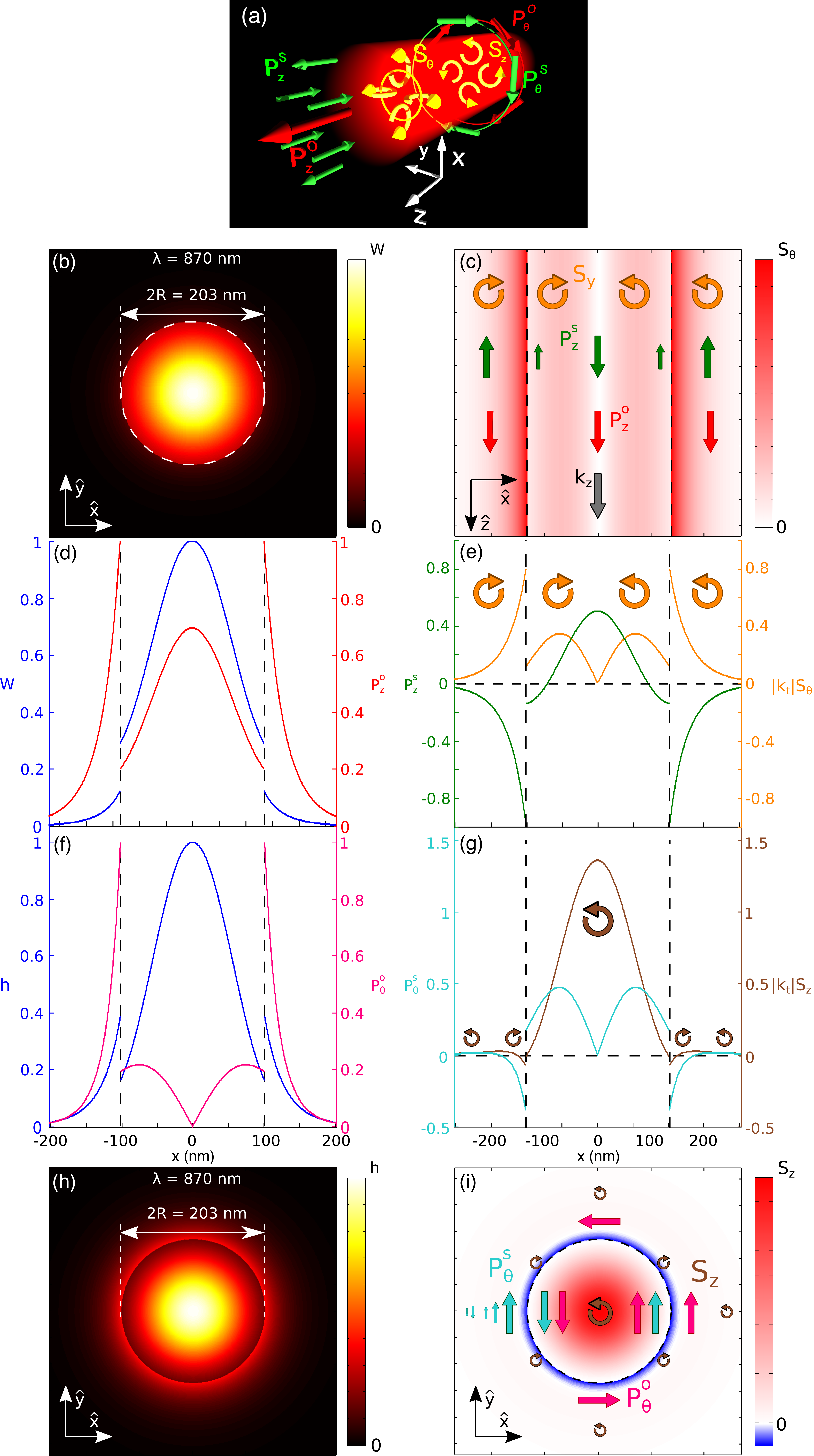}
\caption{(a) Schematic depicting the orientation of orbital (red arrows) and  spin (green arrows) momenta inside the cylindrical 
waveguides; loops illustrate the transverse (respectively, longitudinal) spin densities induced inside the waveguide by the confined nature (respectively, 
helicity) of the hybrid guided modes. The lowest-order hybrid HE$_{11}$ guided mode for a cylindrical waveguide of radius $R$ and $\epsilon_r=11.76$ 
is considered for a normalized radius $\omega R/c=0.73$ (which corresponds to an InP nanowire with $D=2R=230$ nm at $\lambda=870$ nm): 
(b,d) Contour map and radial dependence of the energy density $W$, the latter including also the orbital $P_z^O$ momentum. 
(c,e) Contour map and radial dependence of the only non-zero component of the spin density $S_{\theta}$, 
the latter (e) multiplied by the transverse wavevector component and including also the only non-zero component 
of the spin angular momentum $P_z^S$ (normalized by $P_z^O$).
(f,h) Radial dependence and contour map and of the  helicity density $h$, including also in (f) the helicity-induced transverse 
orbital momentum $P^O_{\theta}$.
(g,i) Radial dependence and contour map in the transverse plane of the helicity-induced longitudinal spin density $S_z$, the former (g)
multiplied by the transverse wavevector component, and including also the related transverse spin angular momentum $P_{\theta}^S$ 
(normalized by $P_{\theta}^O$); dark loops and cyan/magenta arrows indicate spin rotation and transverse spin/orbital momenta.
}
\label{fig_NW_HE}
\end{figure}

Slight differences arise  depending on whether the guided modes is weakly or strongly confined.  What can we expect for the transverse spin from Eq.~(\ref{eq_SNW_TX})? 
Actually, it  depends  linearly on the transverse component $k_t$ of the wavevector inside. Recall that this component is smaller the weaker the confinement is. 
Nonetheless, the energy density $W$, which increases inside the waveguide with increasing confinement, compensates such decrease in a non trivial manner, so that
a compromise between transverse wavevector and energy confinement yields the optimum transverse spin; as an example, we show in Fig. S2 the same
results as in Fig.~\ref{fig_NW_TX}, but for a TE$_{01}$ guided  mode that is weakly confined. Transverse spin density inside the waveguide in Fig. S2 is indeed comparable to that 
in Fig.~\ref{fig_NW_TX}; differences are in turn more obvious  in the transverse spin in the evanescent region, wherein the expected behavior is observed  \cite{Bliokh2014a}: 
more spread outside the cylinder in the latter case (Fig. S2), but larger  close to the waveguide boundary in the former (strongly confined case, see Fig.~\ref{fig_NW_TX}). 
For the sake of completeness, a TM$_{01}$ guided  mode is also shown in Fig. S3; apart from the (relevant) fact that the spin density is entirely electric ($S=S_e$), the qualitative
behavior is very similar to that of the TE$_{01}$ guided  mode  shown in Fig.~\ref{fig_NW_TX}.

Higher-order transverse modes (not explicitly shown here), as for planar guided modes, essentially preserve the (in this case, rotational) symmetry and 
polarization of the spin-related magnitudes, introducing additional lobes in the radial dependence in accordance with the radial mode order $l$
(see e.g.  Fig.~\ref{fig_GHz}  below). Nonetheless, this could lead to various rings alternating  spin rotation inside the cylindrical waveguide,  preserving for all the modes 
the continuity across the cylinder boundary of the outermost ring (even if very thin) with respect to the evanescent-region spin, which remains unaltered.

In order to shed light on the different contributions to the spin density and angular momentum for hybrid modes, we now plot them making special 
emphasis on separating intrinsic-helicity terms from transverse confinement effects. We show all of them in Fig.~\ref{fig_NW_HE}
for a HE$_{11}$ guided mode  in a cross section of the cylindrical waveguide with $\epsilon_r=11.76$,  along with 3D schematics illustrating the 
orientation of the corresponding vectors; the transverse confinement-induced spin (loops) and related momenta (green arrows) are shown 
in Fig.~\ref{fig_NW_HE}c,e, whereas the longitudinal helicity-induced spin and momenta are shown in Fig.~\ref{fig_NW_HE}g,i. 
A color map  of the energy density $W$ revealing mode confinement is included in Fig.~\ref{fig_NW_HE}b, with its radial dependence and that of the orbital 
momentum explicitly shown in Fig.~\ref{fig_NW_HE}d;  the corresponding intrinsic helicity $h$ is included in a contour map in Fig.~\ref{fig_NW_HE}h  for the sake 
of comparison, showing also explicitly its radial dependence in Fig.~\ref{fig_NW_HE}c, along with that of the helicity-induced transverse orbital momentum $P^O_{\theta}$.

First,  a strong transverse (respectively, longitudinal) contribution to the spin density (respectively, extraordinary longitudinal  momentum) is observed
inside, which stems from the transverse confinement ($k_t$-related) as above. Note that the corresponding SAM inside is also concentrated inside an inner ring; 
however, unlike for transverse modes, it rotates similarly to the evanescent-region SAM, and its longitudinal spin  momentum points towards the OAM inside 
(being opposite outside, as expected for evanescent waves). This behavior reveals a richer phenomenology for the confinement-induced transverse SAM governed by 
guided mode symmetry. Second, it is clear that the intrinsic helicity governs the emergence of a strong longitudinal SAM, which leads to an  azimuthal  spin 
momentum density, both resembling the spatial pattern of the helicity density (see Fig.~\ref{fig_NW_HE}h). Its behavior does not differ much from the 
helicity-induced spin exhibited by circularly polarized plane and/or evanescent waves, although again guided mode symmetry renders its phenomenology much richer.
Higher-order hybrid guided  modes (HE$_{ml}$) make the analysis more complex. The rotational symmetry of the all spin-related magnitudes (not shown here) is 
preserved, introducing however alternating spin-sign rings in the radial dependence when the radial index is $l>1$, similarly to higher-order transverse modes 
mentioned above. The azimuthal index $m$ is directly connected to the helicity [cf. Eq.~(S6)], 
as expected; in addition, for $m\geq 2$ the dependence on Bessel functions of the electromagnetic fields inside imposes $W=0$ at the waveguide center.
\begin{figure}
\includegraphics[width=0.9\columnwidth]{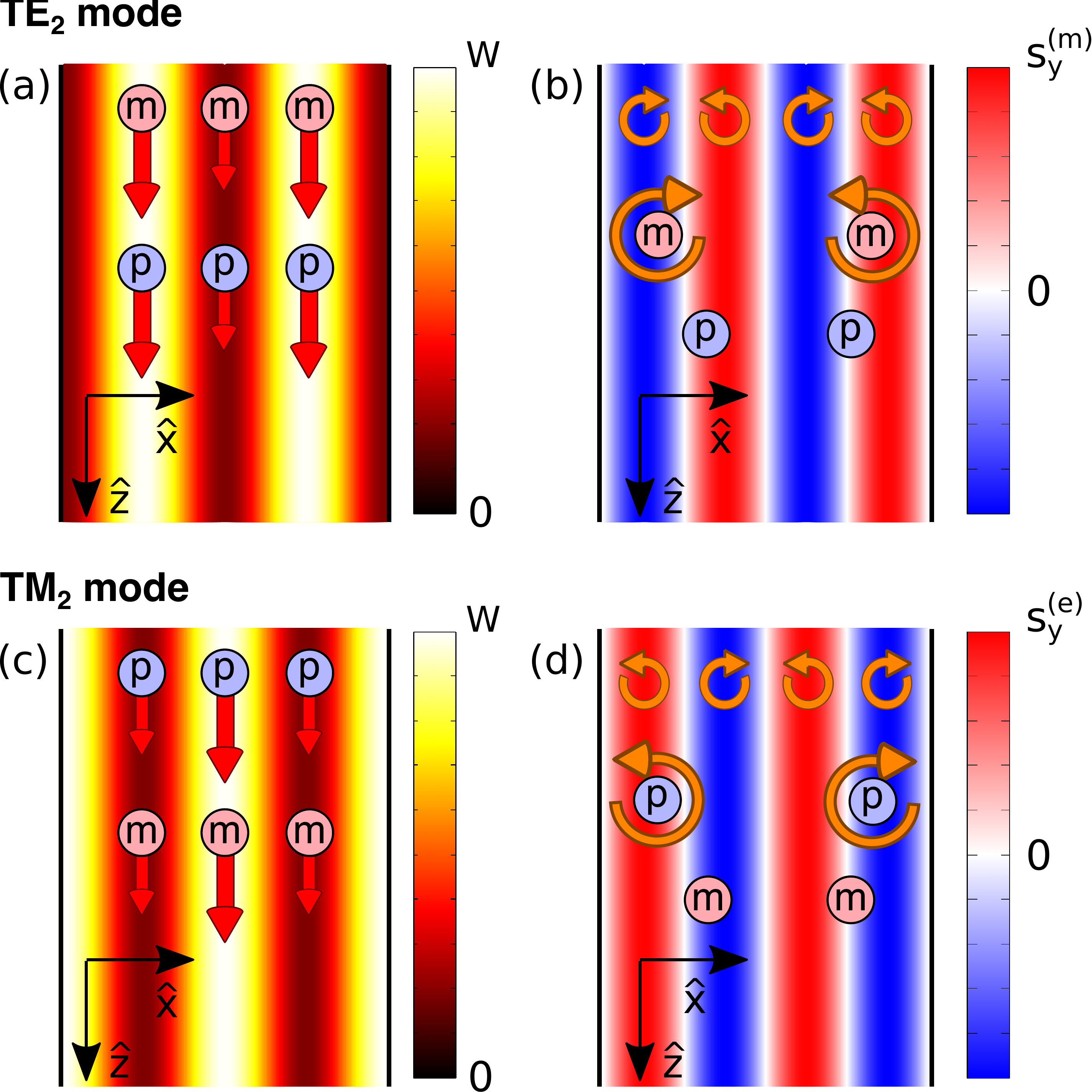}
\caption{(a,c) Energy densities $W$ (left column) and (b,d) spin densities (right column) for 
anti-symmetric TE$_{2}$ (top) and TM$_{2}$ (bottom) modes in a planar waveguide of width $2d$ and $\epsilon_r=80$ 
for normalized  half-width $\omega d/c=0.36$ (which corresponds to water-filled slabs with thicknesses 
$2d=3.4$ cm at $\nu=\omega/(2\pi)=1$ GHz). 
Arrows in (a,c) indicate the radiation pressure force felt by electric ($p$) and magnetic ($m$) dipole particles due to longitudinal momenta, whereas
loops in (b,d) reveal the corresponding torque induced inside the waveguides by transverse confinement-induced spin densities.
}
\label{fig_GHz}
\end{figure}

\section{Optical radiation forces and torques in water-filled waveguides}

Finally, let us briefly explore the impact on optical forces of the transverse SAM inside waveguides shown above. As an example, we
consider water-filled planar channels  (inside e.g. metal plates) in the GHz domain; recall that water in this regime exhibits a large refractive 
index with relatively low absorption, which makes it suitable for tunable high-index-dielectric metamaterials \cite{Andryieuski2015}.
Actually, this scenario could  also be realized for higher frequencies up to the near-IR regime, which is in turn suitable for
optical trapping and manipulation of microscopic particles through optical tweezers \cite{Parkin2007,Nieto-Vesperinas2010,Rodrigues2017}.
The lower-refractive index of water (or other liquids) in this electromagnetic regime comes only at the expense of requiring thicker waveguides.

By way of example, we consider  a planar metallic waveguide filled with water ($n=\sqrt{80}$ at 1 GHz):  particularly, we focus on the asymmetric TE$_{2}$ and TM$_{2}$ 
guided modes  in planar waveguides with $2d=3.4$ cm in Fig.~\ref{fig_GHz}. Assuming that the metallic boundaries behave as a perfect electric conductor,
the dispersion relation reduces to (upon imposing that the transverse electric field at the interfaces vanishes):
\begin{align}
k_xd=(2m+1)\frac{\pi}{2} \;\; \mathrm{or} \;\; k_xd=m\pi  \;\;  (m=0,1,2\ldots).
\end{align}
We would like to stress the fact that, as pointed out above, qualitatively (and nearly quantitatively) similar results would be obtained for a metallic, 
water-filled waveguide with widths of the order of a micron operating in the IR ($\lambda\sim 1.55\mu$m). Along with energy and spin densities, radiation forces and torques felt 
by electric ($p$) and magnetic ($m$) dipole particles stemming from, respectively, longitudinal momenta and transverse SAM, are explicitly revealed through 
arrows and loops. 

First, the rich phenomenology inside with alternating layers/rings with opposite torques is evident  in Fig.~\ref{fig_GHz}b,d. Moreover, 
recall that the SAM may stem from either the electric ($S_e$,  Fig.~\ref{fig_GHz}d) or magnetic ($S_m$,  Fig.~\ref{fig_GHz}b) contribution depending on the nature of the  transverse mode (TM and TE, respectively). 
This implies that $p$ and $m$ dipoles will respond differently to such SAM, exhibiting non-negligible torques only if matching character \cite{Nieto-Vesperinas2010,Picardi2017}:
Thus TE  (respectively, TM) guided modes exert torque only to magnetic (respectively, electric) dipolar particles,  see Fig.~\ref{fig_GHz}b (respectively, Fig.~\ref{fig_GHz}d). 
On the other hand, both electric/magnetic dipolar particles would suffer a radiation force along the guided mode propagation direction, being locally higher at layers with 
larger energy density (see Fig.~\ref{fig_GHz}a,c). Such radiation force $F\sim P^O$ can be anomalously large (respectively, small), as compared to the Poynting vector 
($\mathbf{P}=\mathbf{P^O}+\mathbf{P^S})$, in alternating layers within the waveguide where the resulting spin momentum $P^S$ points opposite, $P^O=P+P^S$, 
(respectively, along, $P^O=P-P^S$), as observed in Fig.~\ref{fig_GHz}a,c.

\begin{figure}
\includegraphics[width=0.9\columnwidth]{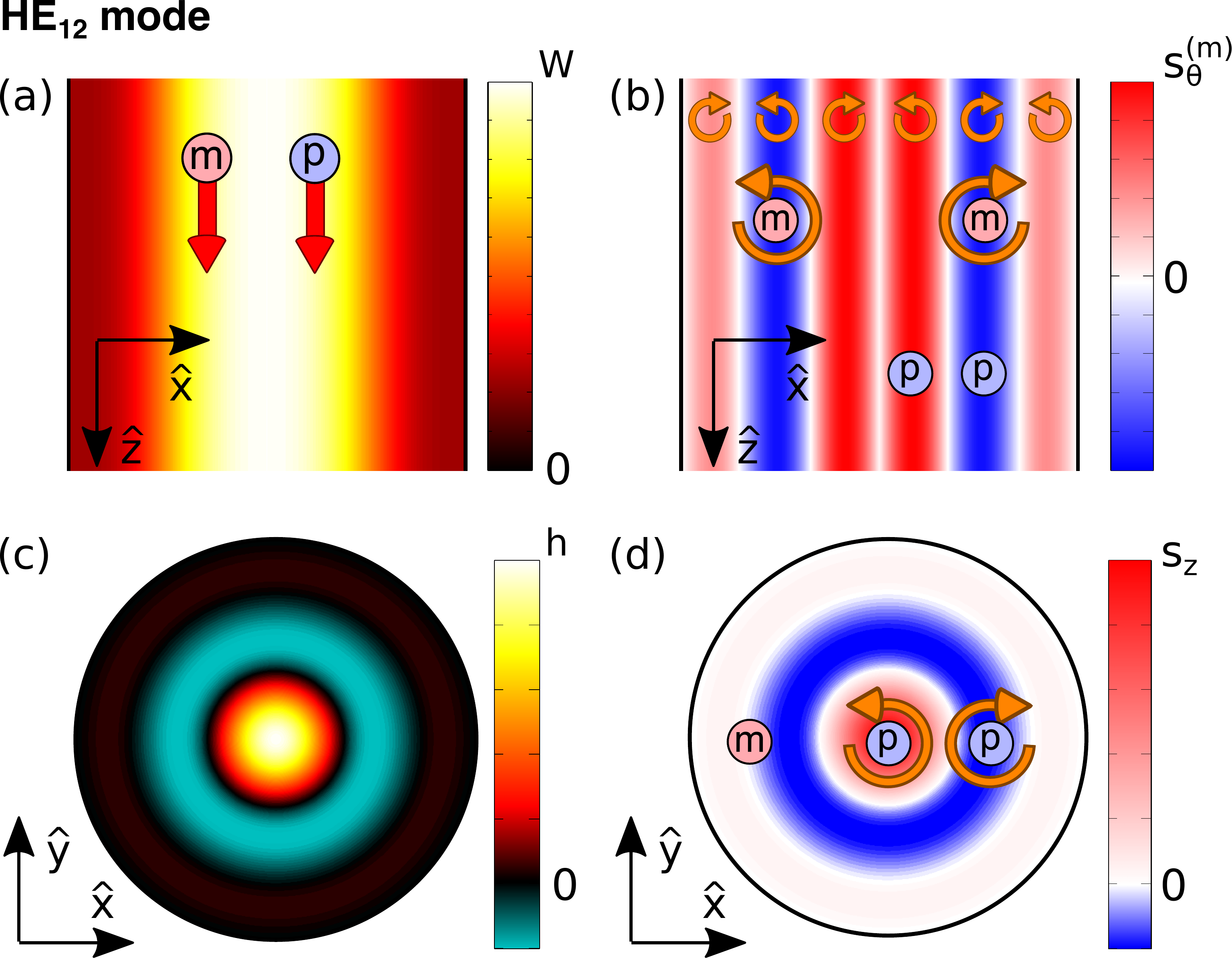}
\caption{(a,c) Energy and helicity densities $W$ (left column) and (b,d) spin densities (right column) for 
 the hybrid  HE$_{12}$ guided mode for a cylindrical waveguide of radius $R$ and $\epsilon^{(r)}=80$  is considered for a normalized radius 
 $R/\lambda=0.097$ (which corresponds to a water-filled cylinder with thickness $2R=5.8$ cm at $\nu=1$ GHz).
Arrows in (a) indicate the radiation pressure force felt by electric ($p$) and magnetic ($m$) dipole particles due to longitudinal momenta, whereas
loops in (b,d) reveal the corresponding torque induced inside the waveguides by spin densities: (b) transverse confinement-induced and (d)
longitudinal helicity-induced.
}
\label{fig_NW_GHz}
\end{figure}

Lastly, we analyze the role of the optical forces on hybrid modes that carried intrinsic helicity. We focus on the HE$_{12}$ hybrid mode of a cylindrical metallic 
waveguide, again, filled by water. In cylindrical waveguides with perfectly conducting walls, guided modes dispersion relation reduces to the values
of $k_tR$ for which the corresponding Bessel functions vanish \cite{Snyder1983}. In particular, the HE$_{12}$ hybrid mode exhibits a varied phenomenology.
First, radiation forces are shown as expected for both probe particles in Fig.~\ref{fig_NW_GHz}a; as above, recall that radiation forces might  be anomalously
larger/shorter than those expected from the Poynting vector density for longitudinal spin momentum opposite/along the canonical momentum. 
Next, the transverse spin density is plotted in  Fig.~\ref{fig_NW_GHz}b, revealing three rings with alternating spin sign inside with vanishing spin density 
at the center. Nonetheless, despite being hybrid, such mode becomes transverse electric so that the spin density contribution is purely magnetic ($S^{(e)}_{\theta}\equiv 0$), 
leading to radiation torque acting only on the magnetic dipole particle $m$. Finally, the helicity density and related spin density are shown in 
Fig.~\ref{fig_NW_GHz}c,d, exhibiting three rings  alternating sign. It should be mentioned that such helicity-dependent spin density has both
electric and magnetic contribution; however, the electric contribution is much larger $S_z^{(e)}\gg S_z^{(m)}$, and  is thus indicated in Fig.~\ref{fig_NW_GHz}d with 
torque only  exerted on the  $p$ probe particle. Interestingly, this implies that an electric (respectively, magnetic) dipole particle would undergo in such a 
waveguide a longitudinal (respectively, transverse) torque. Incidentally,  we have omitted  in Fig.~\ref{fig_NW_GHz} the transverse force induced by the 
longitudinal spin momentum, which actually exerts no radiation force in the dipole approximation,  but does produce a helicity-dependent transverse force  in 
multipolar  interactions with larger  particles \cite{Bliokh2014a}.

It should be emphasized that sub-wavelength particles with high-refractive index have been shown to yield strong magnetic dipole resonances \cite{Garcia-Etxarri2011}, 
as experimentally demonstrated in the optical and lower-frequency domains \cite{Evlyukhin2012,Geffrin2012}, leading to a wealth of phenomenology associated to the 
high-refractive-index dielectric resonant nanostructures \cite{Kuznetsov2016} and sub-wavelength structures in lower frequency regimes in general \cite{Geffrin2012},
where very large refractive indices are indeed ubiquitous.   Recall also that, apart from its impact on induced torques, it has been recently demonstrated that 
such electric/magnetic spin contributions can be experimentally discerned  through high-dielectric-index nanoparticles exhibiting both electric and magnetic dipole 
resonances \cite{Neugebauer2018}. Therefore, multi-resonant sub-wavelength particles would feel electric or magnetic torque depending on the resonant wavelength 
and guided mode involved (the latter influencing also the spatial dependence of such torque), which overall allows for a rich phenomenology.  Finally, recall 
also a related spin-orbit locking is evidently expected if suitable electric or magnetic dipole sources are located inside, which could be exploited all along the 
electromagnetic spectrum.

\section{Concluding remarks}

To summarize, we have analytically investigated the spin and orbital angular momenta of light guided in planar and cylindrical  
waveguides supporting transverse electric/magnetic and hybrid (only in cylinders) modes. Leaving aside the well known transverse spin associated 
to the evanescent component  of the guided modes outside the waveguides, we put the emphasis on the impact of mode confinement inside.
We show that all (transverse an hybrid) modes, despite not being evanescent inside, exhibit a transverse SAM $S_t\propto 
(W/k_z\omega) k_t$ proportional to the 
transverse momentum $k_t$ for planar and cylindrical waveguides; the latter has been attributed the role of an effective mass for guided light 
(dispersion relation such that $\omega/c\sim\sqrt{k_z^2+k_t^2}\sim\sqrt{p^2+m^2}$). Such transverse spin density is shown to carry so called (Belinfante's) longitudinal 
spin angular momentum, parallel or anti-parallel to the proper orbital angular momentum, governed by the mode (spatial) transverse dependence 
imposed by its order and polarization.  Moreover, it is demonstrated analytically that the spin momentum $P^S$ is crucial through this relationship
$P=P^O+ P^S$ ($P$ being related to the Poynting vector momentum density) to retrieve the proper dependence of orbital momentum on propagation 
wavevector $P^O\sim k_z$. This confinement-induced spin is shown in various specific cases to be comparable or larger to that stemming from evanescent 
fields. Furthermore, it exhibits a much richer phenomenology, with either electric or magnetic spin rotating in different directions inside waveguides, 
and even with layers/rings of alternating rotation inside, depending basically on guided-mode spatial and polarization symmetry. Indeed, the 
extraordinary longitudinal spin momentum $P^S$ may also point along or opposite to the orbital momentum inside the waveguide. 
Finally, apart from the former transverse spin, hybrid modes in cylindrical waveguides are shown to carry longitudinal spin arising from its 
intrinsic helicity; such longitudinal SAM concentrates especially inside (strong confinement) the cylindrical waveguide. A discussion is also 
included on the impact of such transverse SAM inside waveguides on optical forces on dipolar particles inside e.g. water-filled waveguides in the IR to
microwave regimes. All this phenomenology has been discussed for lowest-order guided modes in simple planar and cylindrical waveguides for the 
sake of clarity, but can be straightforwardly extrapolated to other geometries preserving guided mode confinement.

Overall, the rich phenomenology demonstrated for such transverse SAM and longitudinal extraordinary momentum associated to guided light opens 
up a wealth of phenomenology for photon spin-orbit coupling inside waveguides beyond what is known for evanescent waves outside. Among the variety 
of configurations where such phenomenology could be exploited and tested, let us mention: 
emitters (quantum dots or wells) located inside semiconductor waveguides/fibers; colloidal quantum-dots in liquid channels; THz or microwave 
electric/magnetic dipole antennas located inside waveguides;  or electric/magnetic dipolar particles for manipulating optical forces inside e.g. tubular, 
liquid-filled micro- or macro-fluidic cavities.

\acknowledgments
This work has been supported by the Spanish Ministerio de Ciencia, Innovaci\'on y Universidades
(LENSBEAM FIS2015-69295-C3-2-P and FPU PhD Fellowship FPU15/03566). 
We are also grateful to M. Nieto-Vesperinas for helpful discussions at the early stages of this work.


%

\begin{widetext}

\largesection
\section{Supplementary material for: Spin angular momentum in planar and cylindrical waveguides induced by transverse confinement and intrinsic helicity of guided light}

\clearpage

\begin{figure}
\includegraphics[width=0.9\columnwidth]{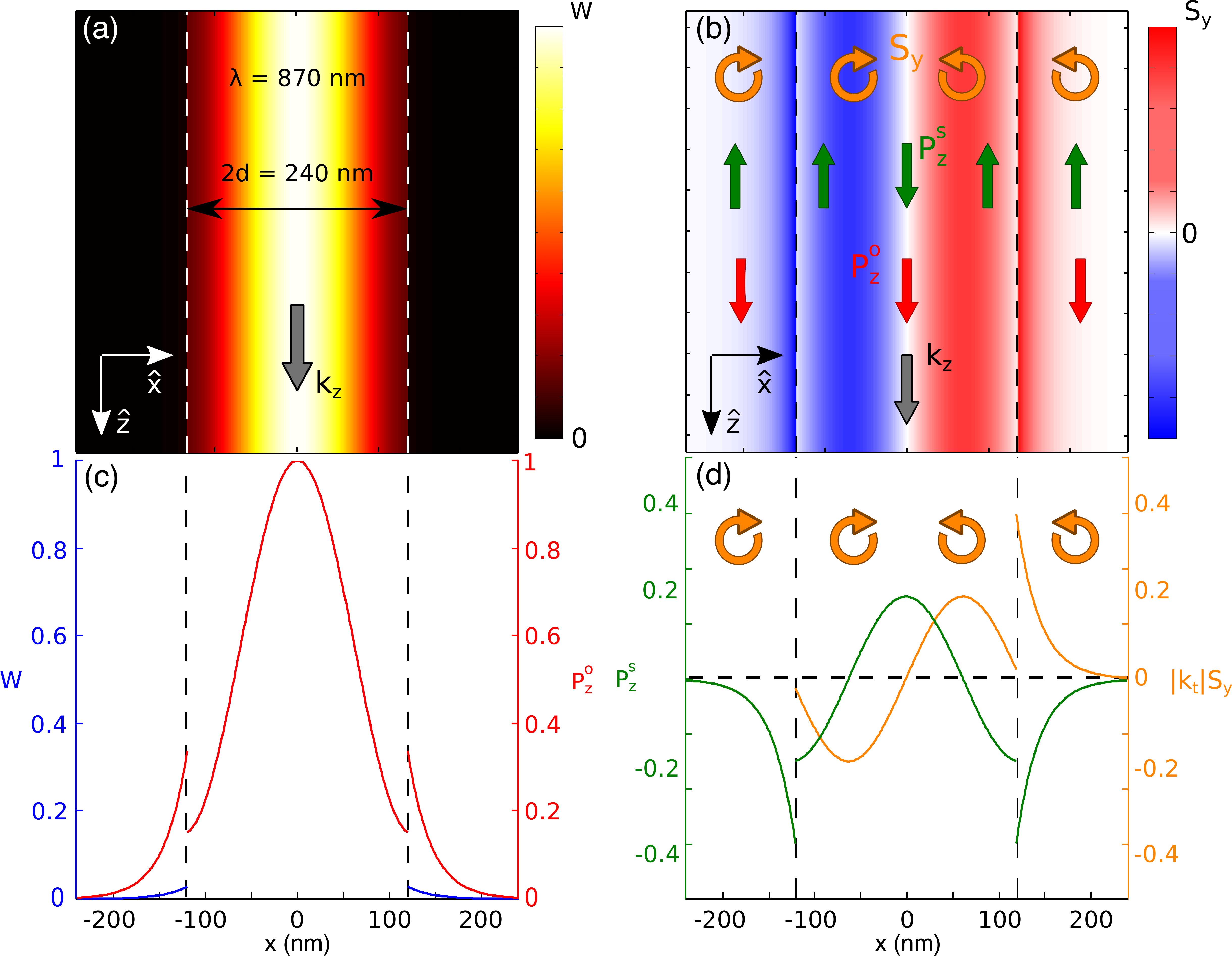}
\caption*{Fig. S1: (a-d) The symmetric lowest-order TM mode is considered
for a planar waveguide of width $2d$ and $\epsilon_r=11.76$ for normalized  half-width $\omega d/c$=0.87 
(which corresponds to an InP nanoslab with thickness $2d=240$ nm at $\lambda=870$ nm).  
(a,c) Contour map and radial dependence of the energy density $W$, the latter including also the orbital $P_z^O$ momentum. 
(b,d) Contour map and radial dependence of the only non-zero component of the spin density $S_y$, 
the latter including also the only non-zero component of the spin angular momentum $P_z^S$.  
}
\end{figure}
\clearpage
\begin{figure}
\includegraphics[width=0.9\columnwidth]{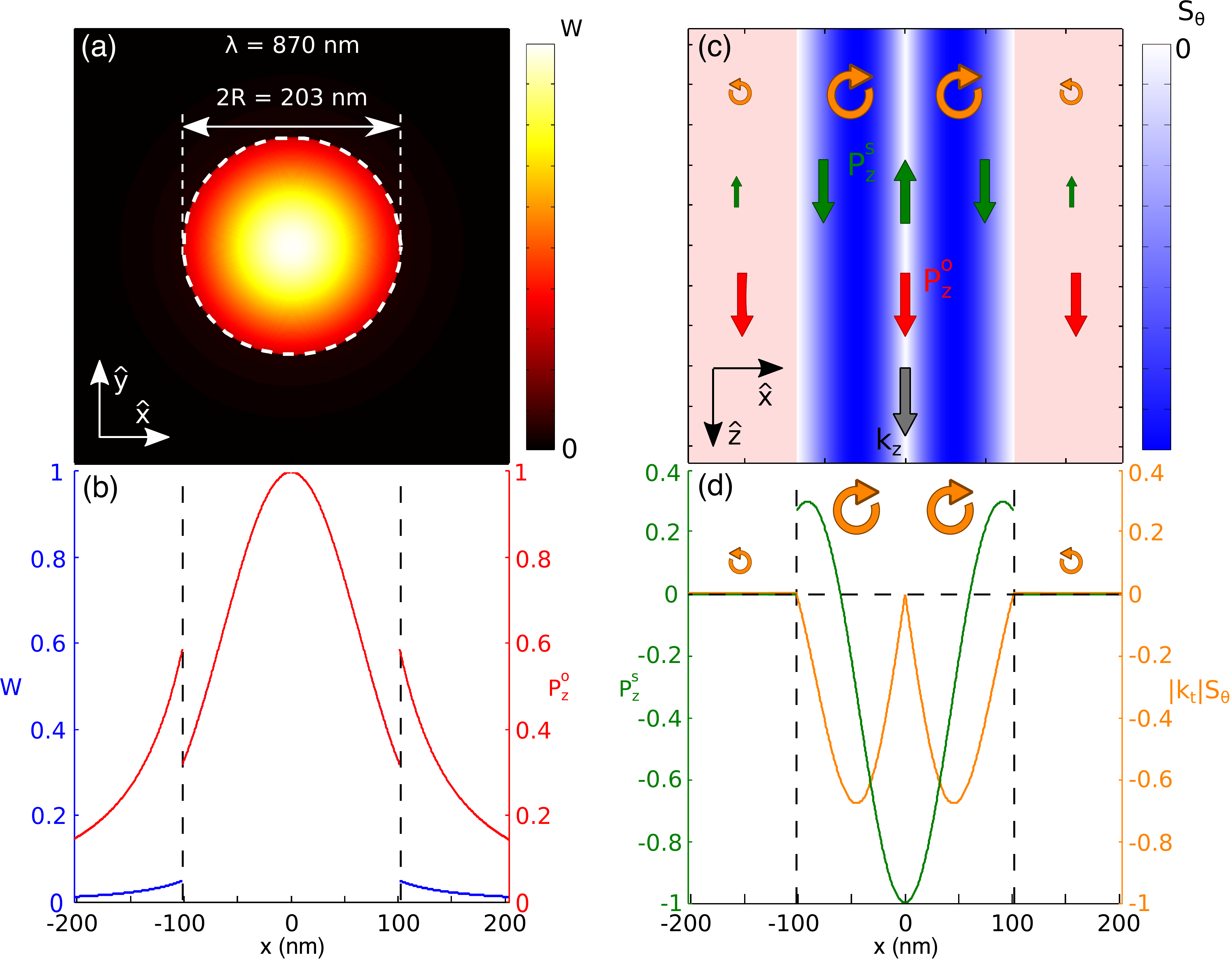}
\caption*{Fig. S2: (a-d) The lowest-order TE$_{01}$ guided mode for a cylindrical waveguide of radius $R$ and $\epsilon_r=11.76$ 
is considered for a normalized radius $\omega R/c=0.73$ (which corresponds to an InP nanowire with $2R=203$ nm at $\lambda=870$ nm, near by the cutoff frequency): 
(a,c) Contour map and radial dependence of the energy density $W$, the latter including also the orbital $P_z^O$ momentum. 
(b,d) Contour map and radial dependence of the only non-zero component of the spin density $S_{\theta}$, the latter including also the 
only non-zero component of the spin angular momentum $P_z^S$.  
}
\end{figure}
\clearpage
\begin{figure}
\includegraphics[width=0.9\columnwidth]{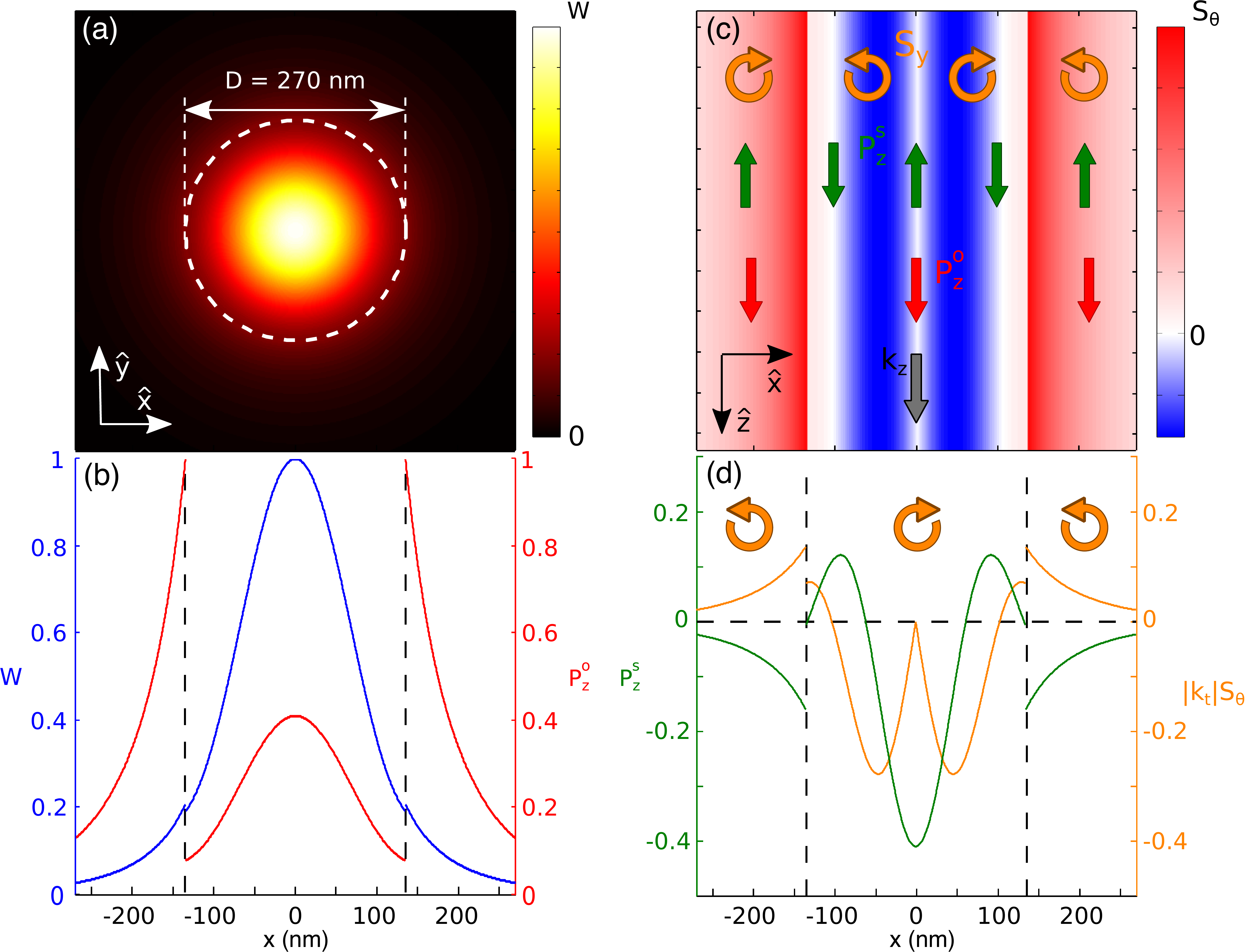}
\caption*{Fig. S3: (a-d) The lowest-order TM$_{01}$ guided mode for a cylindrical waveguide of radius $R$ and $\epsilon_r=11.76$ 
is considered for a normalized radius $\omega R/\lambda=0.97$ (which corresponds to an InP nanowire with $2R=270$ nm at $\lambda=870$ nm): 
(a,c) Contour map and radial dependence of the energy density $W$, the latter including also the orbital $P_z^O$ momentum. 
(b,d) Contour map and radial dependence of the only non-zero component of the spin density $S_{\theta}$, the latter including also the 
only non-zero component of the spin angular momentum $P_z^S$.  
}
\end{figure}
\clearpage

\stdsection
\section{Energy, Poynting vector, and helicity densities in cylindrical waveguides}
\label{app_NW}

The energy density is defined as:
\begin{equation}
W=\frac{\epsilon}{4}|\mathbf{E}|^2+\frac{\mu}{4}|\mathbf{H}|^2.
\tag{S1}
\label{eq_W}
\end{equation}
Upon introducing the electromagnetic fields of guided modes in a cylindrical lossless waveguide given by Eqs.~(20) into Eq.~(\ref{eq_W}),
the resulting total energy density reads:
\begin{align}
&W = \dfrac{1}{4} \left\lbrace \left(|a_{m}|^{2}\epsilon + |b_{m}|^{2}\mu \right)  \left[ \left|Z_{m} \left( k_{t} r \right)\right|^{2} 
+ \dfrac{k_{z}^{2} + k^{2} }{|k_{t} |^{2}}\left(\left|Z_{n}^{'} \left( k_{t} r \right)\right|^{2} + \dfrac{m^{2}}{|k_{t} |^{2} r^{2}} \left|Z_{m} \left( k_{t} r \right)\right|^{2}  \right)    \right] 
\right. \nonumber \\ 
& \hspace{0.9cm}  \left. + \dfrac{8m}{|k_{t} |^{2}r} \dfrac{k_{z}k}{k_{t}}  Z_{m}\left( k_{t} r \right) Z_{m}^{'*} \left( k_{t} r \right)  \Im\left[ a_{m}b_{m}^{*} \right]\sqrt{\epsilon\mu} \right\rbrace.
\tag{S2}
\label{eq_WNW}
\end{align}
The Poynting vector density,
\begin{equation}
\mathbf{P}=\frac{1}{2c^2}\Re\left[\mathbf{E}^*\times\mathbf{H}\right]=\frac{1}{2c^2}\Re\left[\mathbf{E}\times\mathbf{H}^*\right],
\tag{S3}
\label{eq_P}
\end{equation}
yields:
\begin{subequations}\begin{align}
& P_{r} = 0,  \tag{S4a}\\
& P_{\theta} = \dfrac{1}{2\omega} \dfrac{1} {\epsilon_{r} \mu_{r}} \left\lbrace \dfrac{m}{r} \dfrac{k^{2}}{k_{t}^{2}} |Z_{m} \left( k_{t} r \right)|^{2}  
\left( |a_{m}|^{2} \epsilon + |b_{m}|^{2} \mu \right)
+ 2 \dfrac{k_{z} k}{k_{t}} Z_{m}^{'} \left( k_{t} r \right)Z_{m}^{*} \left( k_{t} r \right) \Im\left[a_{m}b_{m}^{*} \right] \sqrt{\epsilon \mu} \right\rbrace, \tag{S4b}\\
& P_{z} = \dfrac{1}{2\omega} \dfrac{1} {\epsilon_{r} \mu_{r}} \left\lbrace\dfrac{k^{2}}{|k_{t} |^{2}} k_{z} \left( |a_{m}|^{2} \epsilon + |b_{m}|^{2} \mu \right) 
\left[|Z_{m}^{'}\left( k_{t} r \right)|^{2} + \dfrac{m^{2}}{|k_{t}^{2}| r^{2}}|Z_{m} \left( k_{t} r \right)|^{2}   \right] 
\right. \nonumber \\[5pt] & \left. 
\hspace{0.9cm} + 2\dfrac{m}{r}\dfrac{k}{k_{t}}\dfrac{k_{z}^2 + k^{2}}{|k_{t} |^{2}} Z_{m} \left( k_{t} r \right)Z_{m}^{'*} \left( k_{t} r \right)  
\Im\left[a_{m}b_{m}^{*} \right]\sqrt{\epsilon\mu}  \right\rbrace.
\tag{S4c}
\end{align}\label{eq_PNW}\end{subequations}
Finally, the helicity can be written as:
\begin{equation}
h=- \frac{\sqrt{\epsilon_{0}\mu_{0}}}{2\omega}\Im[\mathbf{E^*}\cdot\mathbf{H}].
\tag{S5}
\label{eq_h}
\end{equation}
For arbitrary guided mode in a cylindrical waveguide the helicity reads:
\begin{equation}
\arraycolsep=1.4pt\def\arraystretch{1.4}
\begin{array}{ll}
h =& \dfrac{1}{2\omega n} \left\lbrace \dfrac{2m}{|k_{t} |^{2}r} \dfrac{k_{z} k}{k_{t}} \left[|a_{m}|^{2} \epsilon + |b_{m}|^{2} \mu \right] 
Z_{m} \left( k_{t} r \right) Z_{m}^{'*}\left( k_{t} r \right)  
\right.  \\[5pt] &  \left.
+  \left[ \left|Z_{m} \left( k_{t} r \right)\right|^{2} 
+ \dfrac{k_{z}^{2} + k^{2} }{|k_{t} |^{2}}\left(\left|Z_{n}^{'} \left( k_{t} r \right)\right|^{2} + \dfrac{m^{2}}{|k_{t} |^{2} r^{2}} \left|Z_{m} \left( k_{t} r \right)\right|^{2}  \right)    \right]\Im\left[a_{m}b_{m}^{*} \right]\sqrt{\epsilon\mu} \right\rbrace.
\tag{S6}
\label{eq_hNW}
\end{array}
\end{equation}

\section{Spin density and momentum in cylindrical waveguides}
\label{app_NW_s}

The total spin density can be expressed as:
\begin{equation}
\mathbf{S}=\frac{\epsilon_0\mu_r^{-1}}{4\omega}\Im\left[\mathbf{E^*}\times\mathbf{E}\right]+\frac{\mu_0\epsilon_r^{-1}}
{4\omega}\Im\left[\mathbf{H^*}\times\mathbf{H}\right].
\tag{S7}
\label{eq_s}
\end{equation}
For 
arbitrary guided modes in a cylindrical waveguide with electromagnetic field given by Eqs.~(20), the spin density obeys the following expressions:
\begin{subequations}\begin{align}
& S_{r} =  0, \tag{S8a}\\
& S_{\theta}= \dfrac{1}{2\omega}\dfrac{1}{ \epsilon_{r}\mu_{r}}  \left[ \dfrac{k_{z}}{k_{t}}Z_{m}^{*} \left( k_{t} r \right) Z_{m}^{'}\left( k_{t} r \right) 
\left(|a_{m}|^{2}\epsilon + |b_{m}|^{2}\mu \right) 
\right.  \nonumber \\[5pt] & \left. 
\hspace{0.9cm} + \dfrac{2m}{k_{t}^{2}r} k \left|Z_{m} \left( k_{t} r \right)\right|^{2} \Im\left[a_{m}b_{m}^{*} \right] \sqrt{\epsilon\mu} \right],  \tag{S8b}\\[10pt]
& S_{z} =  \dfrac{1}{2\omega} \dfrac{1}{\epsilon_{r}\mu_{r}} \dfrac{1}{|k_{t} |^{2}} \left\lbrace \dfrac{m}{k_{t}r} Z_{m} \left( k_{t} r \right) 
Z_{m}^{'*}\left( k_{t} \right) \left(k_{z}^{2} + k^{2} \right)\left(|a_{m}|^{2}\epsilon + |b_{m}|^{2}\mu \right)  
\right.  \nonumber \\[5pt] & \left. 
\hspace{0.9cm} + 2k_{z} k  \left[ \left|Z_{m}^{'} \left( k_{t} r \right)\right|^{2}  
+ \dfrac{ m^{2}}{|k_{t}|^{2} r^{2}} \left|Z_{m} \left( k_{t} r \right)\right|^{2} \right]\Im\left[a_{m}b_{m}^{*} \right] \sqrt{\epsilon\mu} \right\rbrace,
\tag{S8c}
\end{align}\label{eq_SNW}\end{subequations}
The spin angular momentum,
\begin{equation}
\mathbf{P}^S=\frac{1}{2}\nabla\times\mathbf{S}.
\tag{S9}
\label{eq_ps}
\end{equation} 
 yields:
\begin{subequations}\begin{align}
& P_{r}^{S} =  0, \tag{S10a}\\[5pt]
& P_{\theta}^{S} = \dfrac{1}{4\omega} \dfrac{1}{\epsilon_{r}\mu_{r}} \left\lbrace 
 \dfrac{m}{k_{t}^{2}r}\left[ k_{z}^{2} \left( \left| Z_{m} \left( k_{t} r \right)\right|^{2}  - \dfrac{k_{t}^{2}}{|k_{t} |^{2}}\left( |Z_{m}^{'} \left( k_{t} r \right)|^{2}  + \dfrac{m^{2} }{|k_{t}|^{2} r^{2}} | Z_{m} \left( k_{t} r \right)|^{2} - \dfrac{2}{k_t r}Z_{m} \left( k_{t} r \right) Z_{m}^{'*}\left( k_{t} r \right) \right) \right)  \right. \right. \nonumber \\[5pt] & \left. 
\hspace{0.9cm} + k^{2}\left( (k_t r)^{-1}Z_{m} \left( k_{t} r \right) Z_{m}^{'*}\left( k_{t} r \right) -Z_{m} \left( k_{t} r \right) Z_{m}^{''*}\left( k_{t} r \right) -\dfrac{k_{t}^{2}}{|k_{t}|^{2}}|Z_{m}^{'} \left( k_{t} r \right)|^{2}\right) \right]\left[|a_{m}|^{2}\epsilon + |b_{m}|^{2}\mu \right]  \nonumber\\[5pt] &
\left. 
\hspace{0.9cm}  + 2 k_{z} k \left[ \dfrac{1}{|k_{t} |^{2}r} \left( (m+1)\left|Z_{m}^{'} \left( k_{t} r \right)\right|^{2}  + 2\dfrac{m^{2}}{|k_{t} |^{2} r^{2}} \left|Z_{m} \left( k_{t} r \right)\right|^{2}  \right)   
\right. \right. \nonumber \\[5pt] & \left. \left.
\hspace{0.9cm}  + k_t^{-1}\left(1 - \dfrac{m(3m + 1)}{k_{t}^{2} r^{2}} \right) Z_{m}^{*} \left( k_{t} r \right) Z_{m}^{'}\left( k_{t} r \right)
-k_t^{-1} Z_{m}^{'} \left( k_{t} r \right) Z_{m-1}^{'*}\left( k_{t} r \right)\right]\Im\left[a_{m}b_{m}^{*} \right] \sqrt{\epsilon \mu} \right\rbrace, \tag{S10b} \\[10pt]
& P_{z}^{S} = \dfrac{1}{4\omega} \dfrac{1}{\epsilon_{r}\mu_{r}} \left\lbrace k_{z} \left[ \dfrac{k_{t}^{2}}{|k_{t} |^{2}}\left|Z_{m}^{'} \left( k_{t} r \right)\right|^{2}+ \left(\dfrac{m^{2}}{k_{t}^{2} r^{2}} - 1\right) \left|Z_{m} \left( k_{t} r \right)\right|^{2}\right]\left[|a_{m}|^{2}\epsilon + |b_{m}|^{2}\mu \right]
\right.\nonumber  \\[5pt] &\left. 
\hspace{0.9cm} + \dfrac{4mk}{k_{t}r} Z_{m}^{*} \left( k_{t} r \right) Z_{m}^{'}\left( k_{t} r \right)\Im\left[a_{m}b_{m}^{*} \right] \sqrt{\epsilon \mu} \right\rbrace ,
\tag{S10c}
\end{align}\label{eq_PSNW}\end{subequations}
Finally, the canonical (orbital) part $\mathbf{P}^O$ of the momentum density can be expressed (in cartesians coordinates) as:
\begin{equation}
\mathbf{P}^O=\frac{\epsilon_0\mu_r^{-1}}{4\omega}\Im\left[\mathbf{E^*}\cdot(\nabla)\mathbf{E}\right]+\frac{\mu_0\epsilon_r^{-1}}
{4\omega}\Im\left[\mathbf{H^*}\cdot(\nabla)\mathbf{H}\right],
\tag{S11}
\label{eq_Po}
\end{equation}
where we use the notation
\begin{equation}
(\mathbf{X^*}\cdot(\nabla)\mathbf{Y})_{i} = \sum_{i} X_{i}^{*} \dfrac{\partial}{\partial x_{i}}Y_{j}.
\tag{S12}
\label{eq_ip}
\end{equation}
Bear in mind that Eq.~(\ref{eq_ip}) is only valid in cartesians coordinates. The direct replacement of the $\nabla$ operator by its expression in other coordinates system is wrong. It is necessary to change the expression accordingly.
After some algebraic manipulation we arrive to:
\begin{subequations}\begin{align}
& P_{r}^{O} = 0,  \tag{S13a}\\&
P_{\theta}^{O} = \dfrac{1}{4\omega}\dfrac{1}{\epsilon_{r}\mu_{r}}\dfrac{1}{r} \left\lbrace \dfrac{1}{|k_{t} |^{2}}\left| Z_{m}^{'} \left( k_{t} r \right)  - \dfrac{m}{k_{t} r} Z_{m} \left( k_{t} r \right) \right|^{2} \left[  m\left(k_{z}^{2} + k^{2} \right)\left[|a_{m}|^{2}\epsilon + |b_{m}|^{2}\mu \right] - 4k_{z}k \Im\left[a_{m}b_{m}^{*} \right] \sqrt{\epsilon \mu} \right] 
\right.  \nonumber \\[3pt] & 
\hspace{0.9cm} +\dfrac{2m(m-1)}{|k_{t} |^{2}k_{t} r} Z_{m} \left( k_{t} r \right) Z_{m}^{'*}\left( k_{t} r \right) \left[  \left(k_{z}^{2} + k^{2} \right)\left[|a_{m}|^{2}\epsilon + |b_{m}|^{2}\mu \right] + 4k_{z}k \Im\left[a_{m}b_{m}^{*} \right] \sqrt{\epsilon \mu}  \right]  \nonumber \\[1pt] & \left. 
\hspace{0.9cm} + m\left|Z_{m} \left( k_{t} r \right)\right|^{2} \left[|a_{m}|^{2}\epsilon + |b_{m}|^{2}\mu \right] \right\rbrace, \tag{S13b} \\[10pt]
 & P_{z}^{O} = \dfrac{1}{4\omega} \dfrac{1}{\epsilon_{r}\mu_{r}} k_{z} \left\lbrace\left[ \left|Z_{m} \left( k_{t} r \right)\right|^{2} + \left[ \left|Z_{m}^{'} \left( k_{t} r \right)\right|^{2}  + \dfrac{ m^{2}}{|k_{t} |^{2} r^{2}}\left|Z_{m} \left( k_{t} r \right)\right|^{2} \right] \dfrac{k_{z}^{2} + k^{2}}{|k_{t} |^{2}} \right] \left[|a_{m}|^{2}\epsilon + |b_{m}|^{2}\mu \right]  \right.  \nonumber \\[3pt] &
\left. \hspace{0.9cm} + \dfrac{8m}{|k_{t} |^{2}r} \dfrac{k_{z}k}{k_{t}} Z_{m} \left( k_{t} r \right) Z_{m}^{'*}\left( k_{t} r \right) \Im\left[a_{m}b_{m}^{*} \right]\sqrt{\epsilon \mu} \right\rbrace,
\tag{S13c}
\end{align}\label{eq_poNW}\end{subequations}
which, upon comparing to the energy density (\ref{eq_WNW}), it can be shown that the orbital momentum exhibits the expected dependence with the longitudinal wavevector $k_{z}$: 
\begin{equation}
\dfrac{P^O}{W}=\dfrac{k_{z}}{\omega n^2}. 
\tag{S14}
\end{equation}

\section{Orbital and Spin momenta in cylindrical waveguides for the hybrid HE$_{11}$ mode}
\label{app_NW_HE}

The orbital angular momentum from Eq.~(\ref{eq_poNW}) for the hybrid HE$_{11}$ mode reads: 
\begin{subequations}\begin{align}
&P_{r}^{O} =0 \tag{S15a}\\
& P_{\theta}^{O} = \dfrac{1}{4\omega}\dfrac{1}{\epsilon_{r}\mu_{r}}\dfrac{1}{r} \left\lbrace \dfrac{1}{|k_{t} |^{2}}\left| Z_{1}^{'} \left( k_{t} r \right)  - \dfrac{ Z_{1} \left( k_{t} r \right)}{k_{t} r} \right|^{2} \left[ \left(k_{z}^{2} + k^{2} \right)\left[|a_{1}|^{2}\epsilon + |b_{1}|^{2}\mu \right] - 4k_{z}k \Im\left[a_{1}b_{1}^{*} \right] \sqrt{\epsilon\mu} \right] \right.  \nonumber \\ 
& \left. \hspace{0.9cm} + \left|Z_{1} \left( k_{t} r \right)\right|^{2} \left[|a_{1}|^{2}\epsilon + |b_{1}|^{2}\mu \right] \right\rbrace, \tag{S15b} \\[10pt]
& P_{z}^{O} = \dfrac{1}{4\omega} \dfrac{1}{\epsilon_{r}\mu_{r}} k_{z} \left\lbrace\left[ \left|Z_{1} \left( k_{t} r \right)\right|^{2} + \left[ \left|Z_{1}^{'} \left( k_{t} r \right)\right|^{2}  + \dfrac{ \left|Z_{1} \left( k_{t} r \right)\right|^{2}}{|k_{t} |^{2} r^{2}} \right] \dfrac{k_{z}^{2} + k^{2}}{|k_{t} |^{2}} \right] \left[|a_{1}|^{2}\epsilon + |b_{1}|^{2}\mu \right]  \right. \nonumber \\[5pt]
& \left. \hspace{0.9cm} + \dfrac{8}{|k_{t} |^{2}} k_{z} k \dfrac{Z_{1} \left( k_{t} r \right) Z_{1}^{'*}\left( k_{t} r \right)}{k_{t} r} \Im\left[a_{1}b_{1}^{*} \right]\sqrt{\epsilon\mu} \right\rbrace .
\tag{S15c}
\end{align} \label{eq_PSWNW_HE}\end{subequations}
And the spin momentum, from Eq.~(\ref{eq_PSNW}): 
\begin{subequations}\begin{align}
 & P_{r}^{S} =  0, \tag{S16a} \\
& P_{\theta}^{S} = \dfrac{1}{4\omega} \dfrac{1}{\epsilon_{r}\mu_{r}} \left\lbrace \dfrac{1}{r}\left[ \dfrac{ \left| Z_{1} \left( k_{t} r \right)\right|^{2}}{k_{t}^{2}}  - \dfrac{1}{|k_{t} |^{2}}\left| Z_{1}^{'} \left( k_{t} r \right)  - \dfrac{ Z_{1} \left( k_{t} r \right)}{k_{t} r} \right|^{2} \right]\left(k_{z}^{2} + k^{2} \right)\left[|a_{1}|^{2}\epsilon + |b_{1}|^{2}\mu \right]  \right.  \nonumber \\[2pt]
& \left. \hspace{0.9cm} + 4 k_{z}k \left[ \dfrac{1}{|k_{t} |^{2}r} \left[ \left|Z_{1}^{'} \left( k_{t} r \right)\right|^{2}  + \dfrac{ \left|Z_{1} \left( k_{t} r \right)\right|^{2}}{|k_{t} |^{2} r^{2}}  \right] + \left(1 - \dfrac{2}{k_{t}^{2} r^{2}} \right) \dfrac{Z_{1}^{*} \left( k_{t} r \right) Z_{1}^{'}\left( k_{t} r \right)}{k_{t}}  \right]\Im\left[a_{1}b_{1}^{*} \right] \sqrt{\epsilon\mu} \right\rbrace, \tag{S16b}\\[5pt]
& P_{z}^{S} = \dfrac{1}{4\omega} \dfrac{1}{\epsilon_{r}\mu_{r}} \left\lbrace k_{z} \left[ \dfrac{k_{t}^{2}}{|k_{t} |^{2}}\left|Z_{1}^{'} \left( k_{t} r \right)\right|^{2}  + \left(\dfrac{1}{k_{t}^{2} r^{2}} - 1\right)  \left|Z_{1} \left( k_{t} r \right)\right|^{2}\right]\left[|a_{1}|^{2}\epsilon + |b_{1}|^{2}\mu \right]  \right. \nonumber \\[2pt]
& \left. \hspace{0.9cm} + \dfrac{4}{r} \dfrac{k}{k_{t}} Z_{1}^{*} \left( k_{t} r \right) Z_{1}^{'}\left( k_{t} r \right)\Im\left[a_{1}b_{1}^{*} \right]\sqrt{\epsilon\mu} \right\rbrace . 
\tag{S16c}
 \end{align}\label{eq_PSWNW_HE}\end{subequations}

\end{widetext}

\end{document}